%% file: Main.tex
\documentclass[aps,physrev,preprint,groupedaddress,showkeys,nofootinbib]{revtex4-2}
\usepackage{graphicx}
\usepackage{lmodern}
\usepackage{amsmath}
\usepackage{array}
\usepackage{adjustbox}
\usepackage{multirow}
\usepackage{import}
\usepackage{tikz}
\usepackage{diagbox}
\usepackage{makecell}
\usepackage{calc}
\usepackage{subcaption}
\usepackage{float}
\usepackage{xcolor}
\usepackage[inkscapelatex=false]{svg}
\usepackage{graphicx}
\usepackage{dcolumn}
\usepackage{bm}

\begin{document}

\preprint{APS/123-QED}

\title{Rare-event detection in a backward-facing-step flow using live optical-flow velocimetry: observation of an upstream jet burst }

\author{Juan Pimienta}
\email{ju-pimienta@photonlines.com}
 \affiliation{Photon Lines SAS, Le Black - 10 av. des Touches, 35740 Pacé, France }
\author{Jean-Luc Aider}%
 \email{jean-luc.aider@espci.fr}
\affiliation{PMMH Laboratory, ESPCI, 10 rue vauquelin, 75005 Paris, France}%



\date{\today}

\begin{abstract}
    Rare and extreme events in turbulent flows play a critical role in transport, mixing, and transition, yet are notoriously difficult to capture experimentally. Here we report, to our knowledge, the first direct experimental detection of an upstream-directed jet burst in a backward-facing-step (BFS) flow at $Re_h=2100$, using long-duration Live Optical Flow Velocimetry (L-OFV). Continuous monitoring over 1.5~h enabled a data-driven definition of extremes as rare velocity probes excursions deep into the observed distribution's tails; in practice, large negative events ($u:~Z<-6$, $v:~Z<-5$ at $(x,y)=(2h,h/2)$, where $|Z|>>0$ stands for large deviations from the mean value) triggered the live capture of surrounding velocity fields. The recording is triggered when the probes surpass the defined threshold, using live analysis of the velocity fields. The detected event features a jet-like intrusion into the recirculation region initiated by the collapse of a merged Kelvin–Helmholtz vortex and sustained by counter-rotating vortices, and is accompanied with heavy-tailed probe statistics and simultaneous amplification of fluctuating kinetic energy and enstrophy. While a single event was recorded, underscoring its rarity, the results establish L-OFV as a viable platform for rare-event detection in separated shear layers and document a previously unreported mechanism of upstream jet bursting in BFS flow.
\end{abstract}
\maketitle


\input{Introduction}

\input{Exp_setup}
\input{Detection_protocol}

\input{Event_detected}
\input{Discussion}
\input{conclusions}
\begin{acknowledgments}
This work has been supported by the Association National Recherche Techonologie (ANRT) and Photon Lines SAS. 
\end{acknowledgments}


\bibliography{rare}

\end{document}

%% file: introduction.tex
\section{Introduction\label{sec:intro}}

Rare events can be broadly defined as uncommon fluctuations of a system, often associated with extreme deviations from its typical state. Such events play a crucial role in several disciplines: extreme economic shocks in finance \cite{longin2000value,taleb2007black}, extreme weather events in environmental science\cite{seneviratne2021weather,grant2017evolution} and in epidemiology, rare outbreaks that shape the spread of disease or even pandemics \cite{walters2018modelling, marani2021intensity}. 

In fluid dynamics, rare events also play a critical role in shear flows transition to turbulence. 
A classical example is the subcritical transition from laminar to turbulence, observed in various wall-bounded flows such as Poiseuille \cite{lemoult2012experimental}, Couette \cite{daviaud1992subcritical,duguet2010formation}, Couette-Poiseuille \cite{Klotz2017CouettePoiseuilleFE}, and channel flows \cite{Lemoult2013TurbulentSI,lemoult2014turbulent,Gomé_2021}, where a small change in flow conditions can trigger a violent transition into a chaotic regime. Beyond transition, extreme events also manifest as energetic bursts associated with the breakdown of streaks or hairpin vortices \cite{hack2021extreme}, or in closed turbulent flows where intermittent structures reorganize the system \cite{Moffatt_2021}. At larger scales, rare-event algorithms have shown how turbulent jets can undergo structural transitions driven by the nucleation of coherent vortices\cite{bouchet2019rare}. Similarly, rare or intermittent transitions have been documented in wakes and bluff-body flows, producing large fluctuations in aerodynamic forces\cite{gayout2021rare,KharsanskyAtallah2024FromLO,varon2017chaotic,lestang2020numerical}. From a practical perspective, such extremes are crucial for fluid-structure interactions: engineering designs must endure the most violent loads, not just the mean ones. 

However, capturing and analyzing rare events is not an easy task precisely because of their rarity and unpredictability. A variety of theoretical and numerical approaches have been developed for their detection. \citet{giardina2011simulating} introduced a cloning Monte Carlo method, rooted in large deviation theory to enhance the sampling of rare trajectories in stochastic and deterministic systems. This framework has been applied to extreme hydrodynamic forces exerted on a square cylinder \cite{lestang2020numerical}. However most of these works are numerical studies where extreme events can be accessed and identified, even in computationally demanding cases such as isotropic turbulence \cite{Yeung_2015}. Experimental identification of such events remains a formidable challenge. The rarity of such events means their timing and location are unpredictable, making long-duration monitoring essential \cite{LOHSE_intermittency}.

Most experimental approaches have relied on local probes such as hot wires \cite{Rao1971TheP,Morrison1988ConditionalsamplingSF} or wall pressure or shear sensors \cite{orlu2020instantaneous,Whalley2019AnEI} to detect the occurrence of a rare event. The advantage of such sensors is that they provide high temporal resolution and information can be recorded for a long time.  The drawback is that they are limited to only a few isolated localized regions of the flow and are often either intrusive (hot wire) or limited to the wall region. In contrast, particle image velocimetry (PIV) offers better spatial resolution and gives access to a large region inside the flow, but is not adapted to rare-event detection because of the impossibility to run very long measurements, unless coupled with a local intrusive sensor that triggers the recording like in \cite{Whalley2019AnEI}.

In this work, we propose to overcome these challenges using Live Optical Flow Velocimetry (L-OFV), which combines real-time computation of 2D velocity fields inside the flow, with continuous long-term monitoring. Applied to a canonical separated flow -- the Backward-Facing Step (BFS) flow -- this technique enables us to define statistical thresholds for rare events and to capture the associated flow fields exactly when they occur. We focus here on extreme deviations of local velocity probes as indicators of rare dynamics, and report, to our knowledge, the first experimental detection of a singular event: a strong upstream jet penetrating into the BFS recirculation zone.

The paper is organized as follows. Section \ref{sec:setup} introduces the experimental setup and L-OFV methodology, including the detection protocol. Section \ref{sec:detection_protocol} establishes the statistical criteria for rare-event definition based on long-duration monitoring. Section \ref{sec:event_detected} presents the detected event, with analysis of its dynamics and flow structures. Section \ref{sec:discussion} discusses the statistical signatures and physical interpretation. Finally, Section \ref{sec:conclusions} concludes and outlines perspectives for future work.

%% file: Exp_setup.tex
\section{Experimental setup\label{sec:setup}}

\subsection{Hydrodynamic Channel and BFS geometry\label{subsec:hydro_tunnel}}

\begin{figure}[H]
    \centering
    \includegraphics[width=0.8\linewidth]{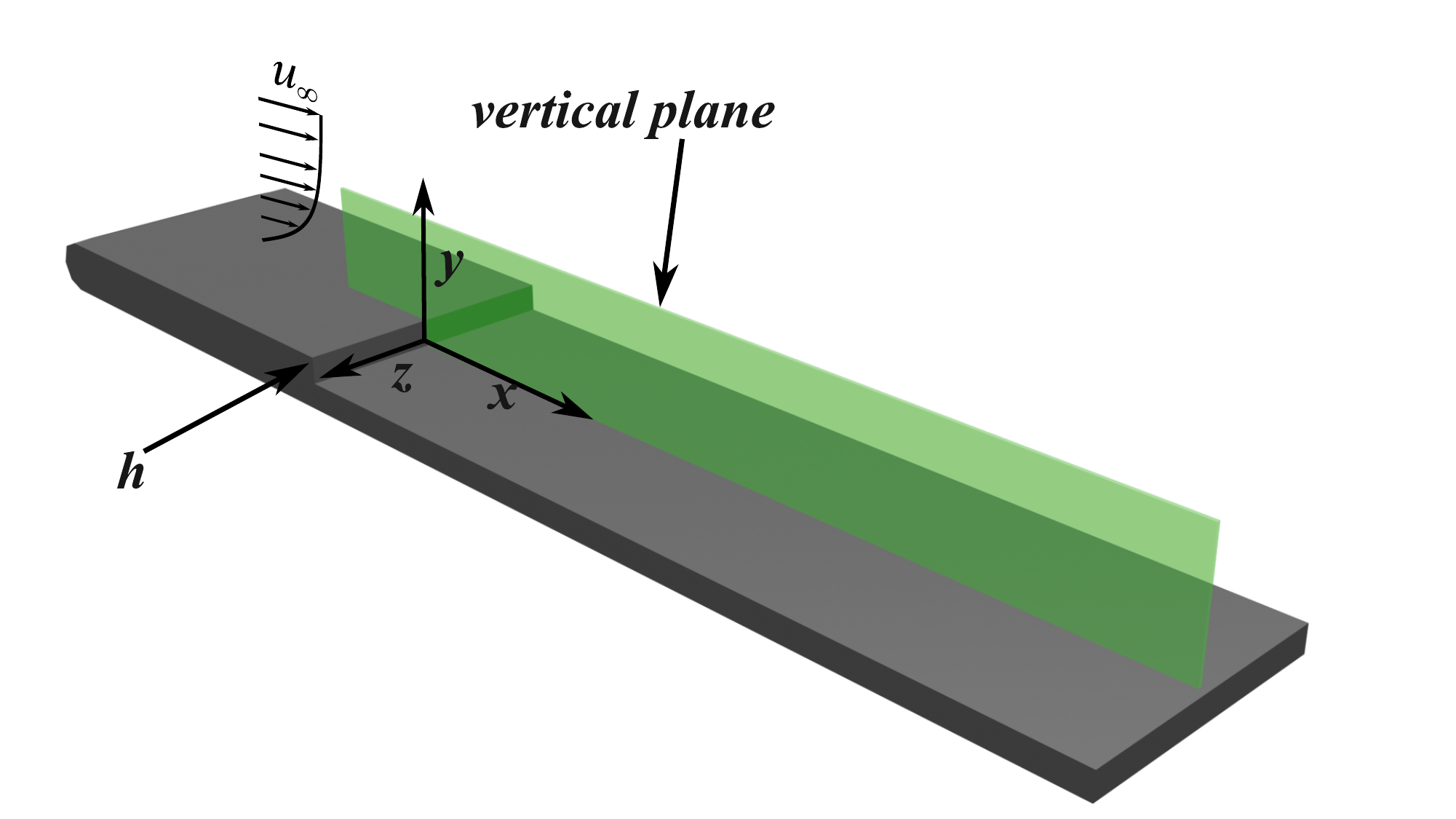}
    \caption{3D sketch of the BFS model used in the experiments, showing the measurement plane located at the center of the yz plane. }
    \label{fig:BFS_3L}
\end{figure}
The  BFS model  consists of a sharp-edged step over a flat plate that generates a massive boundary-layer separation, followed by a reattachment downstream on the lower wall. This a classic geometry to generate a well-defined flow separation. The step edge is located at $x=25~cm$ downstream of the plate's leading edge. The geometry is shown in Fig.~\ref{fig:BFS_3L} which also illustrates the location of the laser sheet used for the PIV measurements. The step height is $h=1.5~cm$ for a channel height $H=7~cm$, leading to a vertical expansion ratio $E=\frac{H+h}{H}=1.21$. The test section is $80~cm$ long and has a rectangular cross-section of width $w~=15~cm$. The origin of the coordinates $(x, y, z) = (0, 0, 0))$ is located at the base of the step, in the middle of the channel. 

The Reynolds number based on the step height is  $Re_h=\frac{U_\infty h}{\nu}$, where $\nu$ is the kinematic viscosity of water and $U_\infty $ is the longitudinal freestream velocity. In our setup, the maximum available free-stream velocity is $U_\infty\approx22~cm/s$, leading to a maximum Reynolds number $Re_h\approx3300$ at room temperature ($20^oC$).

\subsection{Live Particle Image Optical Flow Velocimetry\label{subsec:OFV}}
A Coherent Genesis\texttrademark MX continuous laser ($\lambda=532~nm$, 2 W) provided the illumination. The \textcolor{black}{laser} beam passes through a $60^o$ cylindrical Powell lens to form a $\sim 1.5 ~mm$ thick light sheet, with relatively uniform illumination across the measurement plane. Neutrally buoyant light-reflecting polyamide microparticles ($\phi\approx20~\mu m$) were used to seed the flow. 


The choice of the camera was critical for long-duration, relatively high frequency measurements. The main requirements were a continuous and fast image streaming to a workstation and high acquisition frequencies under continuous illumination. For this purpose, a Mikrotron 21CXP12 camera was used, providing images of $5120\times896~px^2$ ($3.27~Mp$) with an acquisition frequency of 100 Hz. The pixel size is $\approx83.3~\mu m$.

The images were processed in real time using eyeMotion and eyePIV softwares. Eyemotion is the parent image processing software of eyePIV, the optical flow velocimetry (OFV) software developed in collaboration between Photon Lines and the PMMH laboratory. \textcolor{black}{The OFV code is based on the Lucas-Kanade optical flow algorithm \cite{Lucas1981AnII} and was initially optimized to run on Graphics Processing Units (GPU). The CUDA language (CUDA being the programming language developped by Nvidia for their GPUs) allows for real-time computing of the velocity fields because the L-K algorithm is highly parallelizable and adapted to the GPU architecture. The first applications of real-time OFV were as a visual sensor in closed-loop flow control experiments using a LabView environment} \cite{Gautier2013RealtimePF,Gautier2013ControlOT,Gautier2015FrequencylockRC,gautier2015closed,varon2019adaptive}. With the latest software optimizations and one of the latest GPU,  up to thousands of dense velocity fields (one velocity vector per pixel) per second can be computed \textit{live}, depending on image size and OFV parameters. 

In general, the method consists of comparing the intensity gradients between two pixel-centered kernels from two images. The per/pixel displacement is found by an iterative minimization, and the process is repeated through resolution-degraded images to estimate different size of displacements. A pre-processing step of normalization completed the process. There are, hence, four parameters in the algorithm: kernel radius (\emph{KR}), number of iterations (\emph{IT}), pyramid sub-levels (\emph{PSL}), and normalization radius (\emph{NR}). The companion article to this paper presents an in-depth characterization of the algorithm, error analysis, and computational capabilities. For the following experiments the OFV parameters were: $KR=7$ ~px, $NR = 4$, $IT = 4$, $PSL = 3$.

The workstation is a custom-built prototype powered by an AMD Ryzen Threadripper PRO 3955WX CPU (16 cores at 3.9 GHz) with 128 GB RAM, equipped with a last generation Nvidia RTX5090 GPU dedicated to live OFV processing.  

\subsection{\textcolor{black}{Optical Flow Velocimetry measurements precision\label{subsec:OFV-validation}}}

\begin{figure}[H]
    \centering
    \begin{subfigure}{0.3\linewidth}
        \centering
        \includegraphics[width=\linewidth]{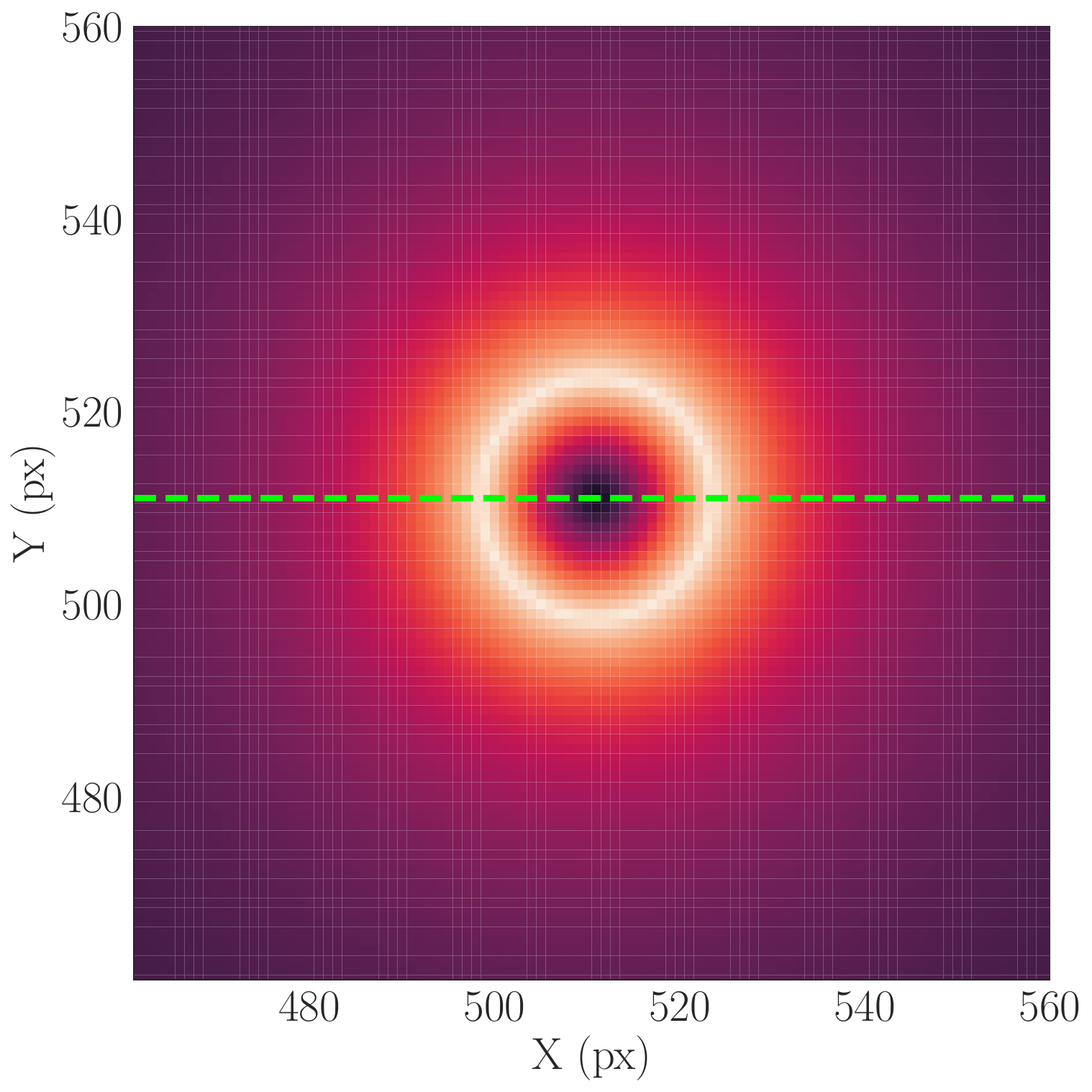}
        \caption{}
        \label{fig:th_r12}
    \end{subfigure}
    \centering
    \begin{subfigure}{0.3\linewidth}
        \centering
        \includegraphics[width=0.985\linewidth]{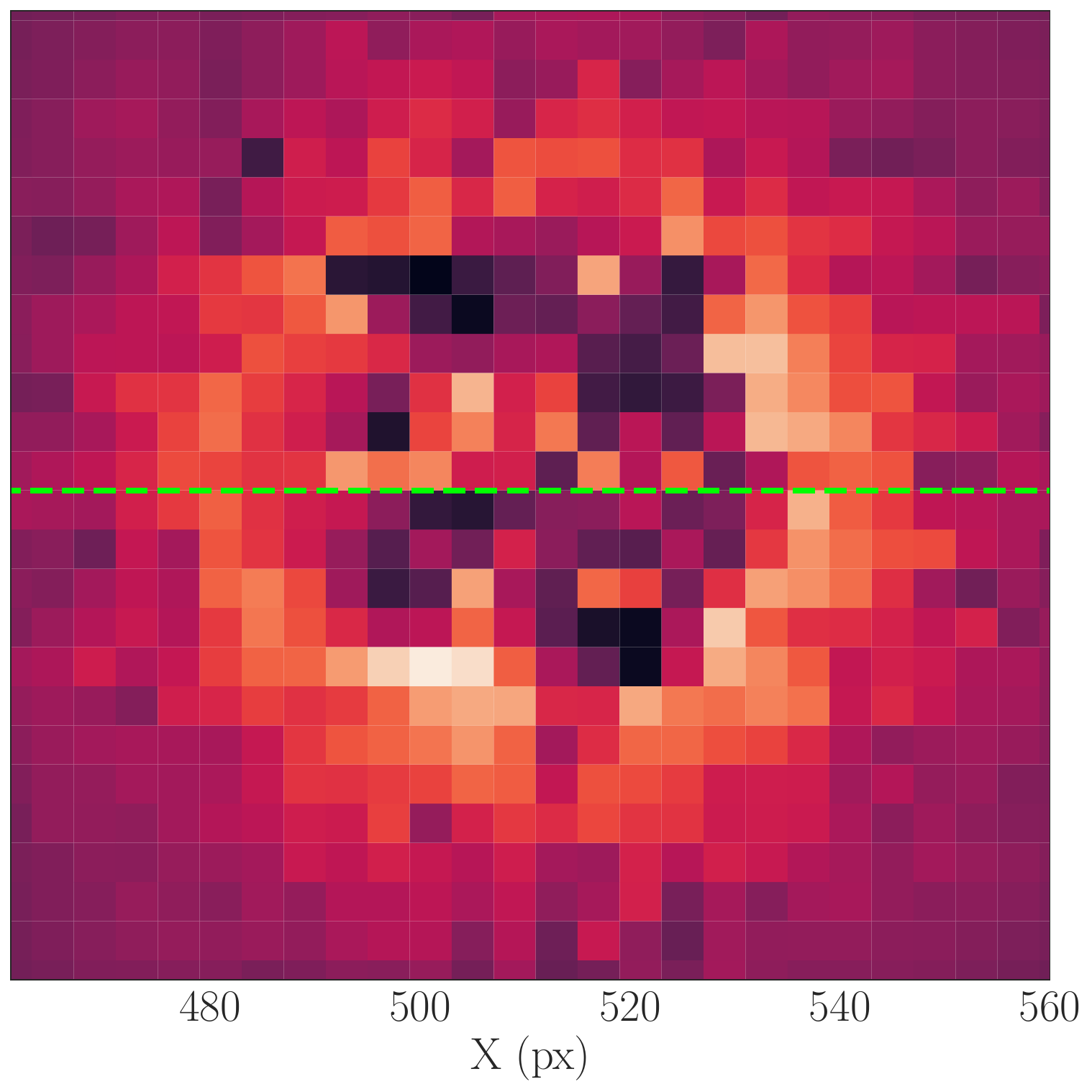}
        \caption{}
        \label{fig:cc_r12}
    \end{subfigure}
    \centering
    \begin{subfigure}{0.3\linewidth}
        \centering
        \includegraphics[width=\linewidth]{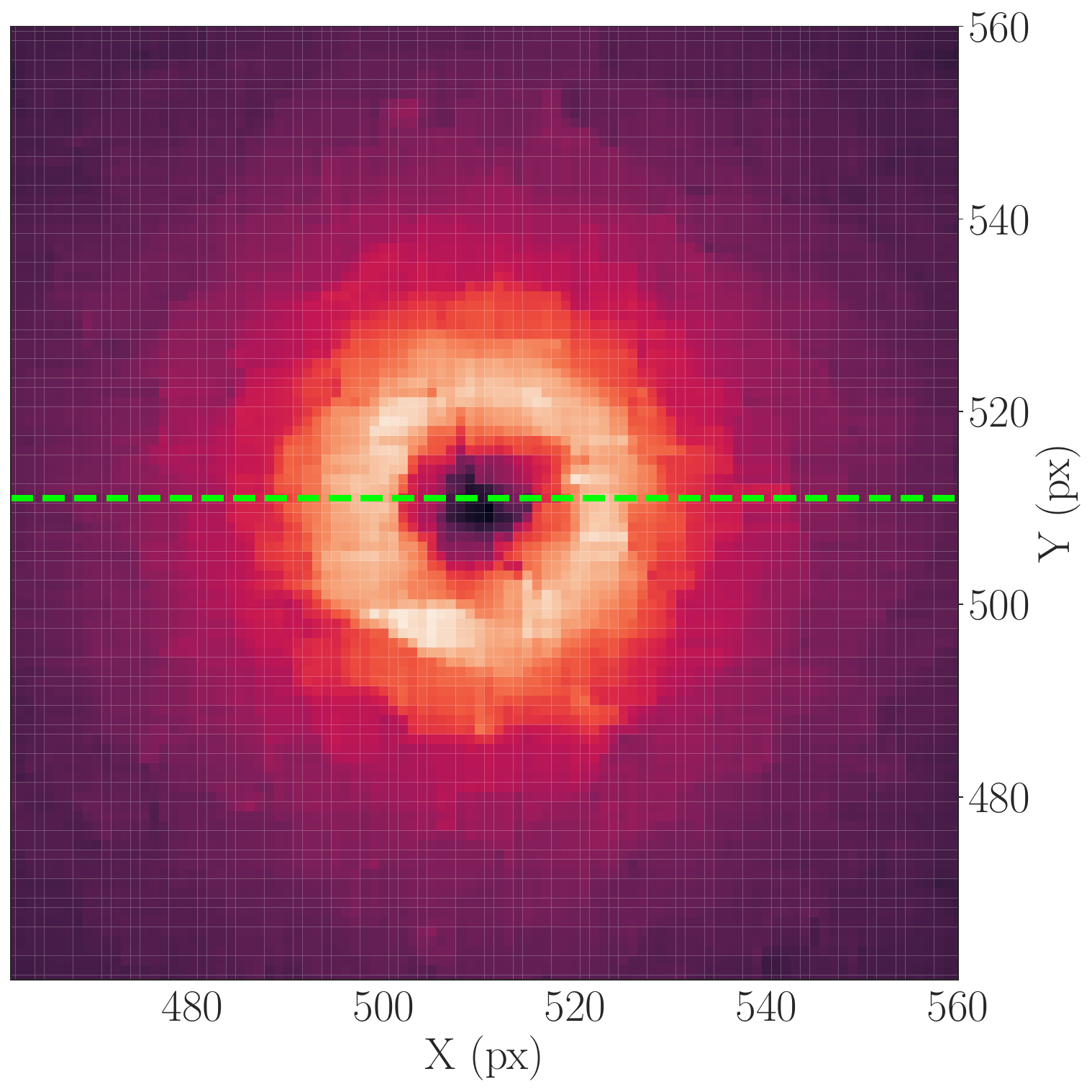}
        \caption{}
        \label{fig:of_r12}
    \end{subfigure}
    \begin{subfigure}{0.9\linewidth}
        \includegraphics[width=\linewidth]{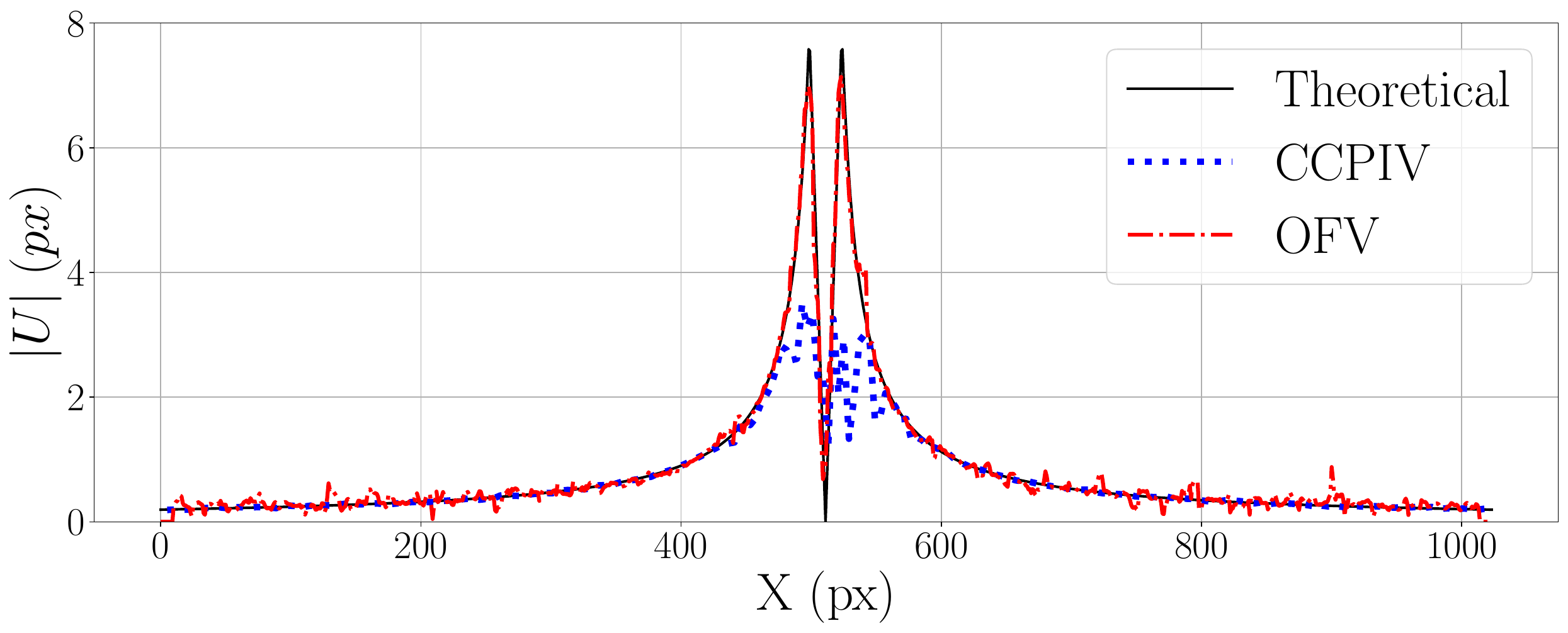}
        \caption{}
        \label{fig:r12_profiles}
    \end{subfigure}
    \caption{\textcolor{black}{Rankine vortex with $r=12$~px and $D=8$~px. \textbf{(a)} Theoretical displacement zoom. (b) CC-PIV zoom. \textbf{(b)} OFV zoom. \textbf{(c)} Mid-height displacement-magnitude profile (dotted green line). OFV from an image pair with 15~p/IW. From \cite{pimienta2025high}.}}
    \label{fig:rk_comparison_12}
\end{figure}

\textcolor{black}{In \citet{pimienta2025high} an error analysis was thoroughly performed on OFV by the use of synthetically generated particle images, using as a test case a Rankine vortex with various vortex core sizes ($R$) and displacements ($D$). This study served to assess OFV's error margins and uncertainty of measurement in challenging situations such as the estimation of small vortices with large displacement. It was found that, if right experimental conditions are met, OFV's error margins are sub--pixel for the most part. A clear comparison of OFV capabilities is presented in Fig.~\ref{fig:rk_comparison_12}, where the results of the displacement field of a small Rankine vortex is shown. The vortex core has a radius $r=12$~px with a displacement of almost 8 px. This combination portrays a very challenging test case of high displacement gradient over a small region in space, as portrayed by the theoretical displacement field in Fig.~\ref{fig:th_r12}. To provide a comparison, Fig.~\ref{fig:cc_r12} shows the result of the displacement field using CC-PIV\cite{Thielicke_2021}, while  Fig.~\ref{fig:of_r12} shows OFV's result, both over a small region of the field where the vortex is located. It is clear that OFV shows a more coherent estimation of such a complicated case, while CC-PIV fails particularly in the center region of the vortex. This is confirmed when comparing displacement profiles at mid-height of the vortex from each of the cases, and comparing them to the analytical solution, shown in panel Fig.~\ref{fig:r12_profiles}. OFV capability to better resolve spatially concentrated displacement gradients is evident. Even though the maximum of displacement are underestimated by $5\%$, OFV manages to locate accurately the position of the displacement peaks.}

\begin{figure}
    \centering
    \includegraphics[width=0.6\linewidth]{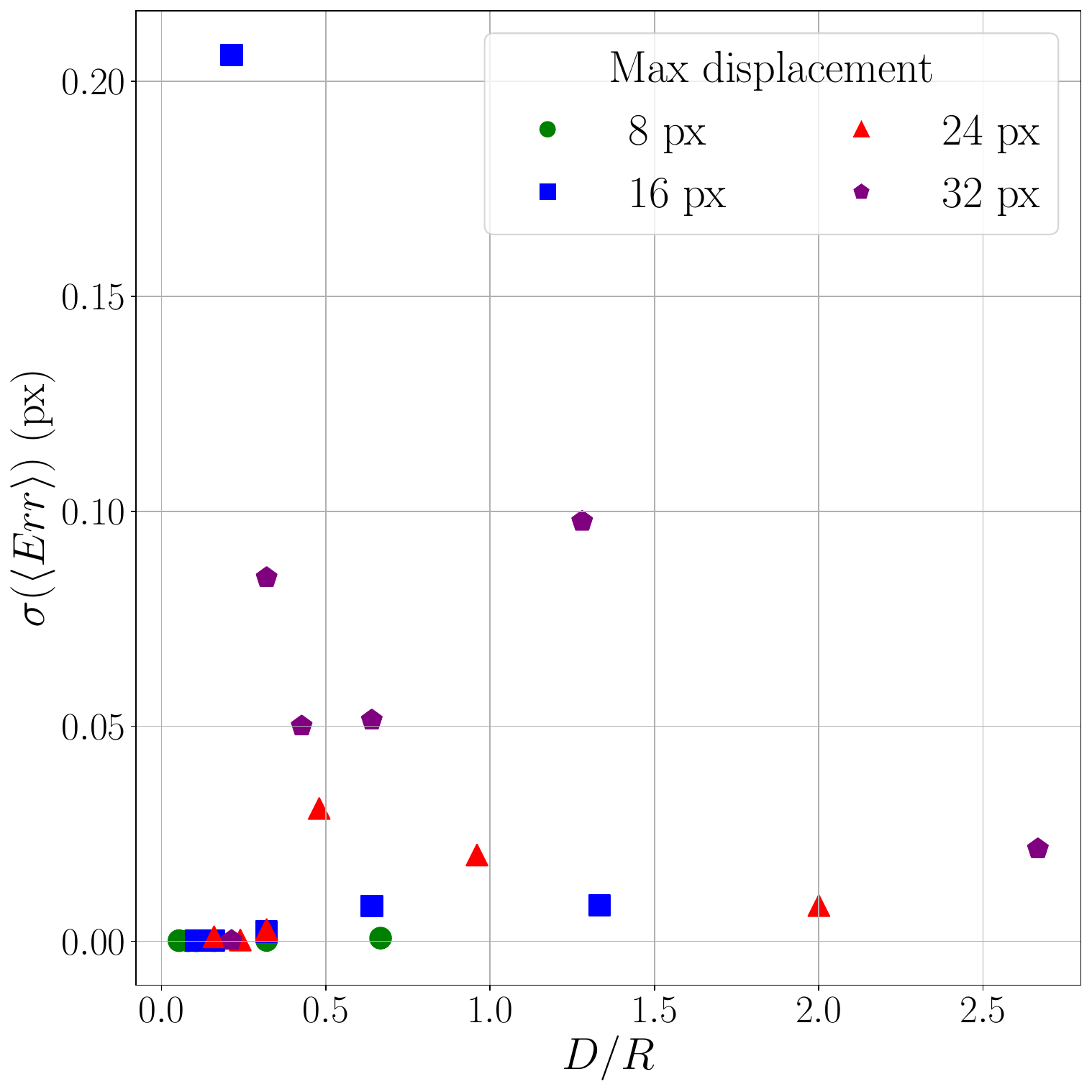}
    \caption{\textcolor{black}{Standard deviation computed over 10 independent synthetic realizations for each $(D,R)$ case using OFV. From \cite{pimienta2025high}.}}
    \label{fig:rk_std}
\end{figure}

\textcolor{black}{The repeatability of this results was also analyzed through sets of 10 individual realizations of the different Rankine cases tested to study the deviation of the results. Fig.~\ref{fig:rk_std} shows the standard deviation of the absolute displacement error between OFV's results and the analytically resolved Rankine case over the 10 different realizations. The results are presented as function of $D/R$ which served as a displacement gradient proxy, and discriminated by the maximum expected displacement $D$ in the legend. The dispersion of the error is low, with majority of the results presenting a standard deviation of the error of the order $\sigma(\langle Err\rangle) \sim O(10^{-4})$~px. Certain cases portray higher orders of variation, however in general the deviation remains sub--pixel, which tells about OFV's displacement estimation reliability. For more details, see \cite{pimienta2025high}.} 


%% file: Detection_protocol.tex
\section{Determining a rare and extreme event in the BFS flow\label{sec:detection_protocol}}

\subsection{BFS flow\label{subsec:BFS_flow}}

\begin{figure}[!h]
    \centering
     \begin{subfigure}{\linewidth}
        \includegraphics[height=0.4\linewidth]{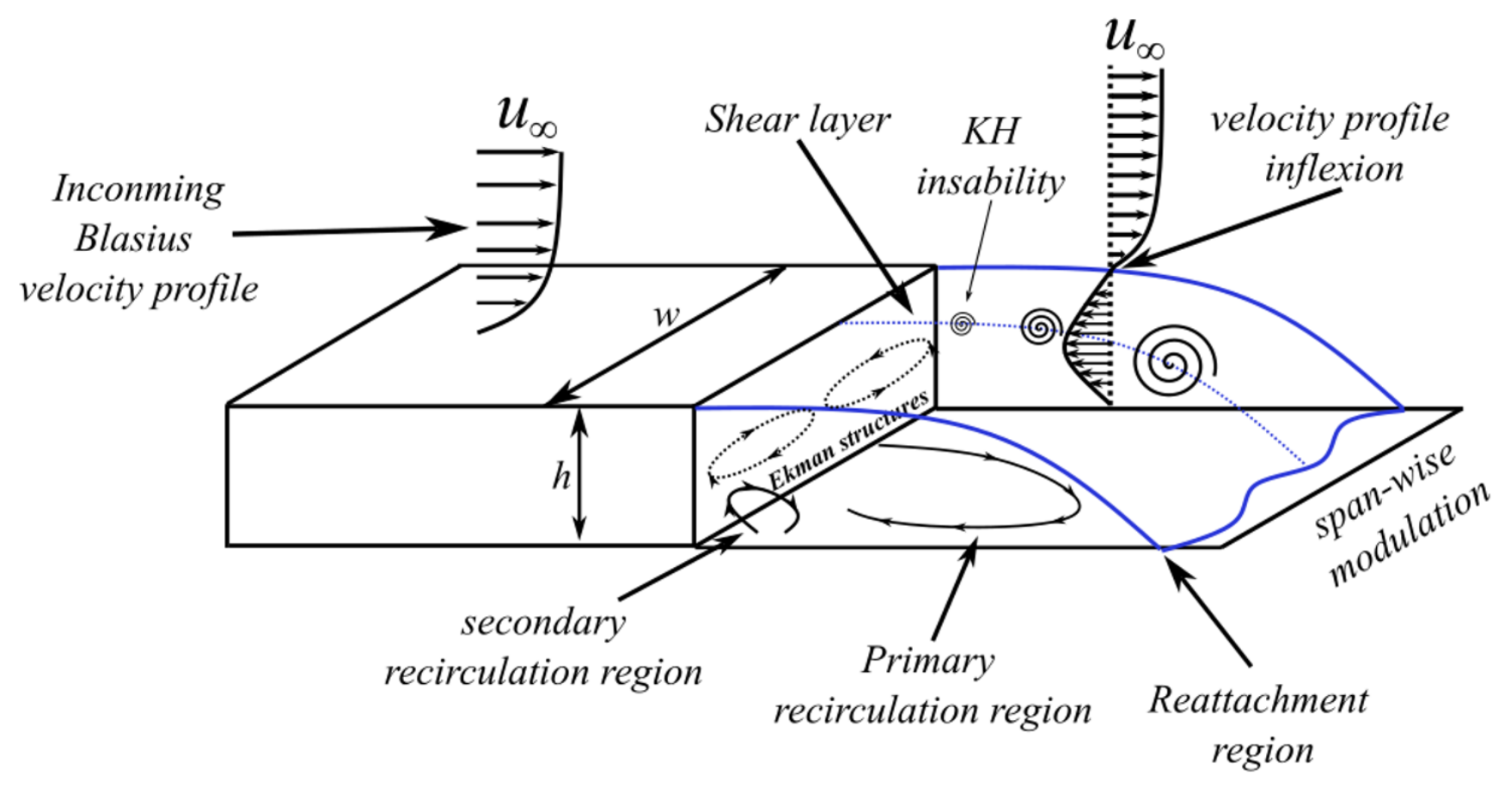}
        \caption{}
    \end{subfigure}
    \begin{subfigure}{\linewidth}
    \includegraphics[width=\linewidth]{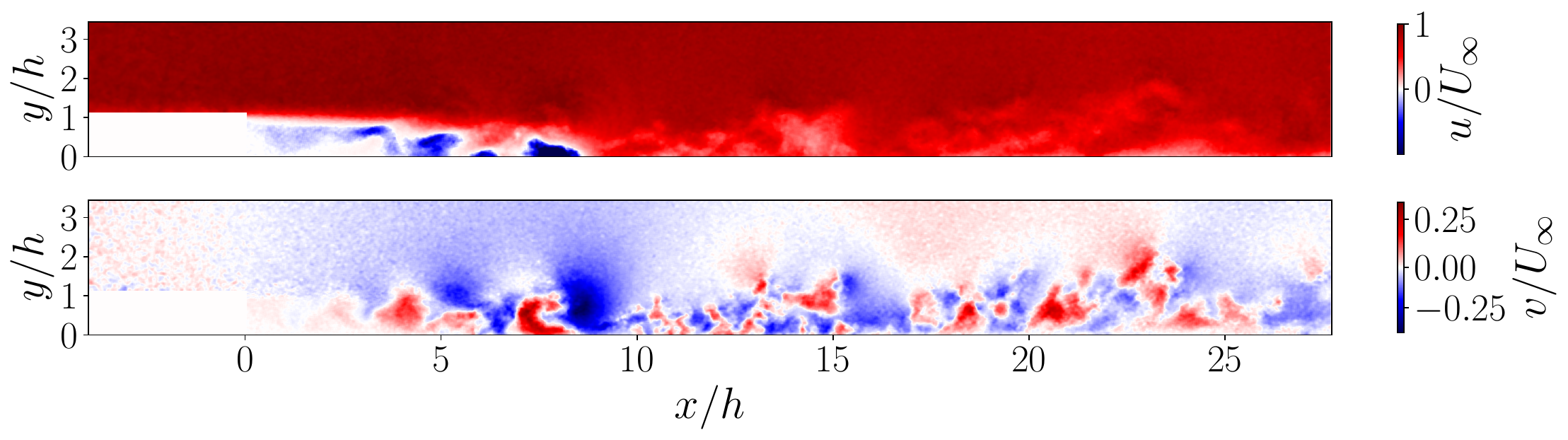}
    \caption{}
    \end{subfigure}
    \caption{ a) Sketch showing the main instabilities and  main structures generated downstream the step edge. b) Instantaneous $u(x, y)$ (upper \textcolor{black}{field}) and $v(x, y)$ (lower \textcolor{black}{field}) velocity fields measured in the vertical symmetry plane ($z/h=0$), downstream of the BFS, at $Re_h=2100$. $x=0$ is taken at the step edge, while $y=0$ is defined at the horizontal wall downstream the step.}
    \label{fig:BFS_flow}
\end{figure}

The BFS flow has been widely studied experimentally but also numerically because of the simplicity of its geometry leading to a very complex 3D flows \cite{armaly1983,le1997direct,kostas2002particle,Aider2007LargeeddySA}. As shown in Fig.~\ref{fig:BFS_flow}~a), the incoming boundary layer separates at the step edge, forming a shear layer which grows downstream. Kelvin-Helmholtz vortices are shed in the curved shear layer, which reaches the lower wall further downstream in the reattachment region. A large recirculation area, also called recirculation bubble, is formed, under the shear layer, characterized by a much slower negative streamwise velocity. Of course, with increasing Reynolds number, all these vortical structures go through various destabilization processes leading to the creation of more and more 3D transverse and streamwise structures \cite{Aider2007LargeeddySA}, ultimately leading to a complex full 3D turbulent flow. An example of transitional flow, measured at $Re_h=2100$ in the vertical symmetry plane of the flow, close to turbulence, is shown in Fig.~\ref{fig:BFS_flow}~b), where the instantaneous streamwise and vertical velocity fields are visualized. The search for rare events will take place in this symmetry plane.

\subsection{Rare events detection protocol\label{subsec:detecion_protocol}}

\begin{figure}[h]
    \centering
    \includegraphics[width=\linewidth]{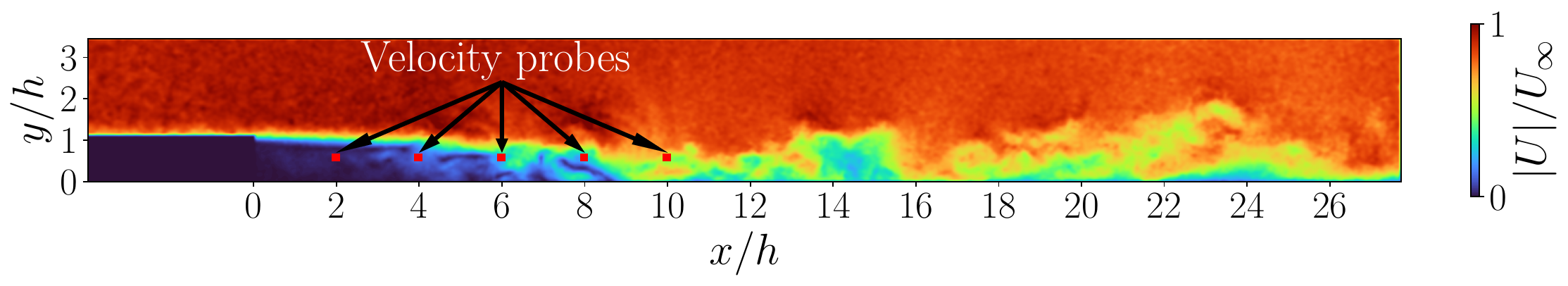}
    \caption{Instantaneous velocity field magnitude in the vertical symmetry plane ($z/h=0$), downstream of the BFS, at $Re_h=2100$. Velocity probes are located at the mid-height of the step ($y/h=0.5$) and at five streamwise . \textcolor{black}{The probes are numbered from 1 ($x/h=2$) to 5 ($x/h=10$).}}
    \label{fig:VF_probes}
\end{figure}

Using our Live-OFV system it is possible to compute at each time-steps various quantities from the instantaneous 2D velocity fields and then to search for a statistics-based protocol for rare event detection. Because a continuous recording of instantaneous 2D velocity fields is not possible over hours,  local probes have been defined inside the 2D instantaneous velocity fields computed in real-time. Various locations were tested. Finally, five pairs of velocity probes $(u,v)$ were defined in the velocity field, as illustrated in Fig.~\ref{fig:VF_probes}. They are located at the mid-height of the step, and of the recirculation bubble, and are distributed regularly along the streamwise direction from $x=2h$ to $x=10h$. These probes are used as local sensors of the flow state. The time series of $(u,v)(t)$ measured for each  probe could then be recorded continuously, as long as needed, enabling statistical definition of rare events without requiring terabytes of image storage. 

As mentioned previously, the full 2D velocity fields could not be recorded indefinitely, even if computed continuously. Instead, a circular memory buffer of 1 000 continuous images was maintained during the full process of detection (i.e. hours). The algorithm was configured to record permanently a full 1 000 images sequence only when an extreme event was detected. Practically, when a probe signal crossed a predefined threshold, the system triggered recording,  keeping 500 frames before and 500 frames after the detection time. This ensured that the full temporal development of the rare event, before and after its detection, was captured.

\subsection{Rare events detection criteria\label{subsec:detecion_protocol}}

To detect rare events in the BFS flow, preliminary studies were needed to define appropriate detection criteria. Long-duration  observations were thus carried out, during which time series of local quantities were measured and recorded.

\begin{figure}[h]
    \begin{subfigure}{\textwidth}
        \includegraphics[width=\linewidth]{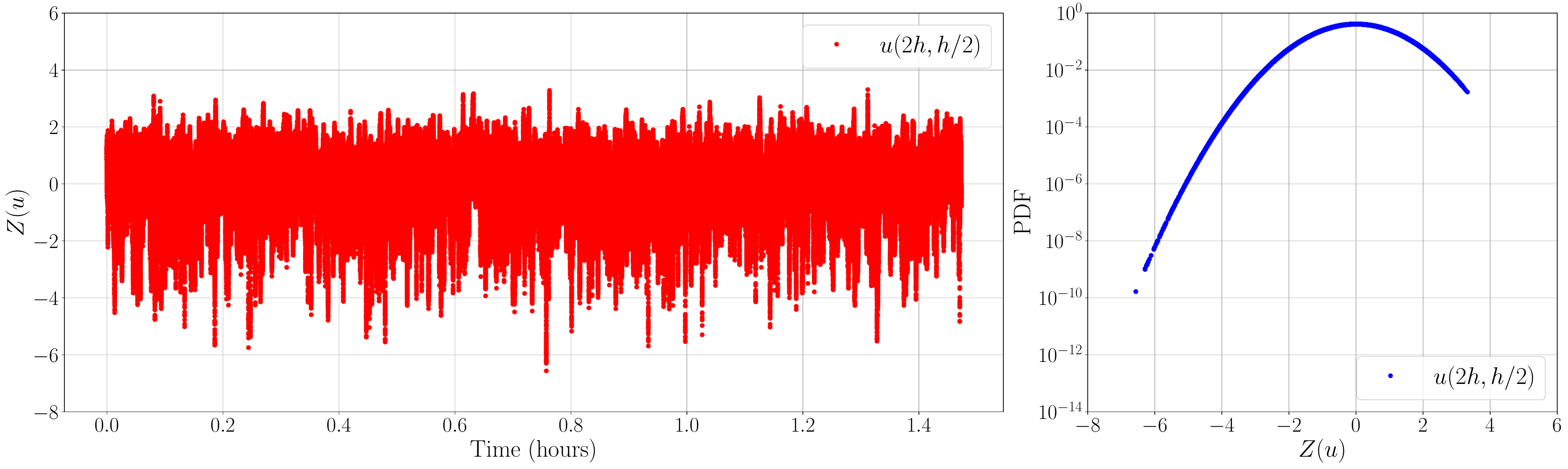} 
        \caption{}
        \label{fig:probes_u_0}
    \end{subfigure}
    \begin{subfigure}{0.48\textwidth}
        \includegraphics[width=\linewidth]{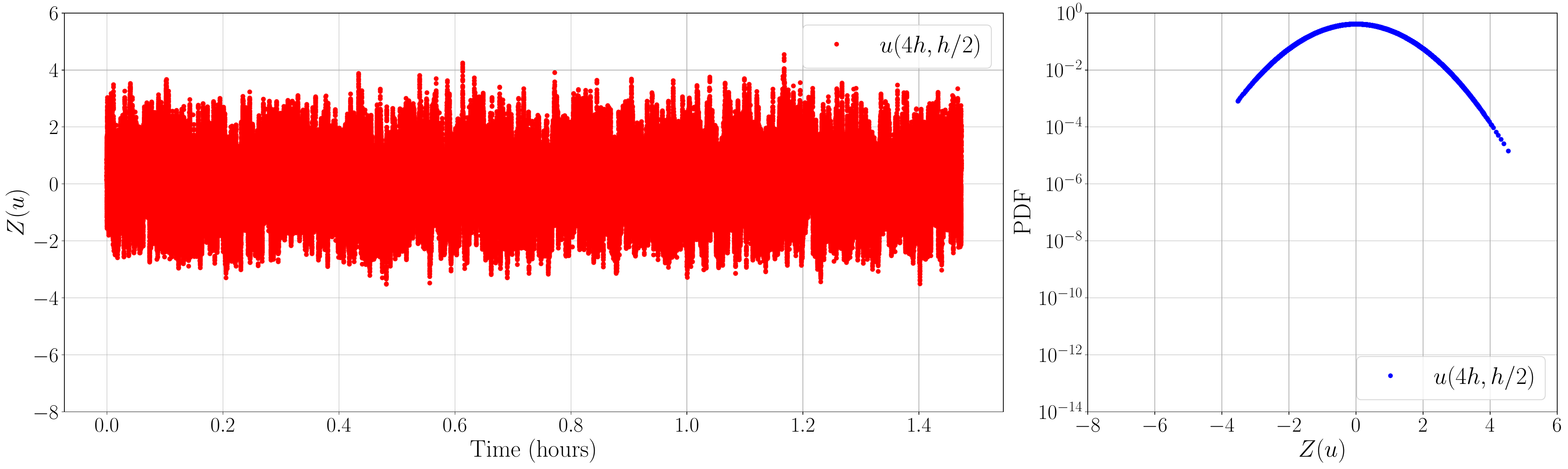}
        \caption{}
        \label{fig:probes_u_1}
    \end{subfigure}   
    \begin{subfigure}{0.48\textwidth}
        \includegraphics[width=\linewidth]{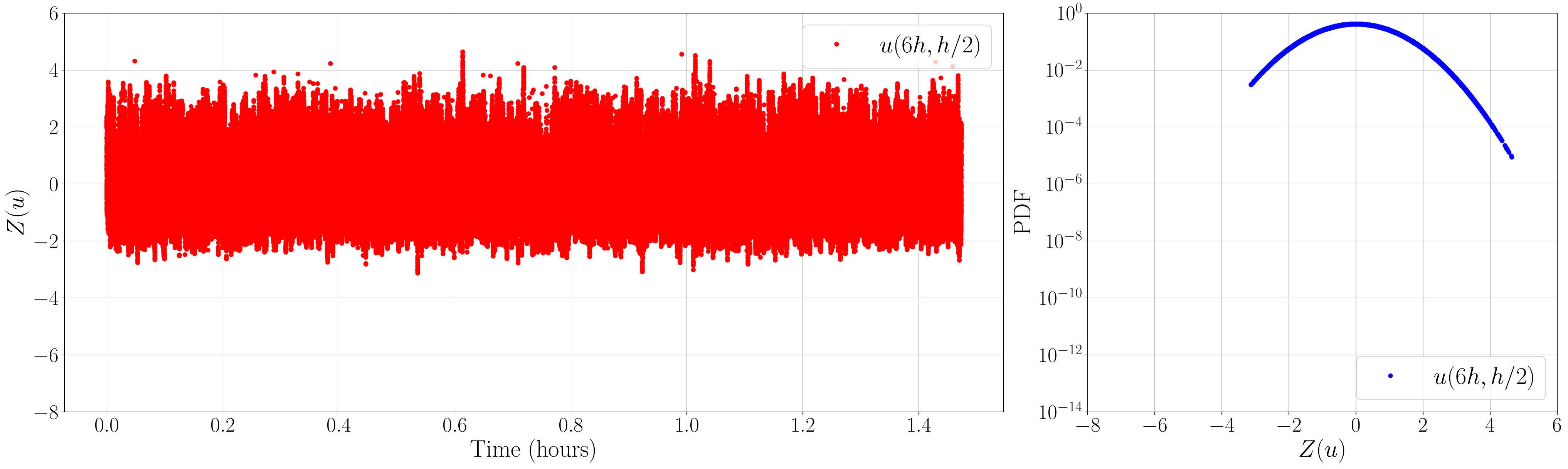}
        \caption{}
        \label{fig:probes_u_2}
    \end{subfigure}      
   \begin{subfigure}{0.48\textwidth}
        \includegraphics[width=\linewidth]{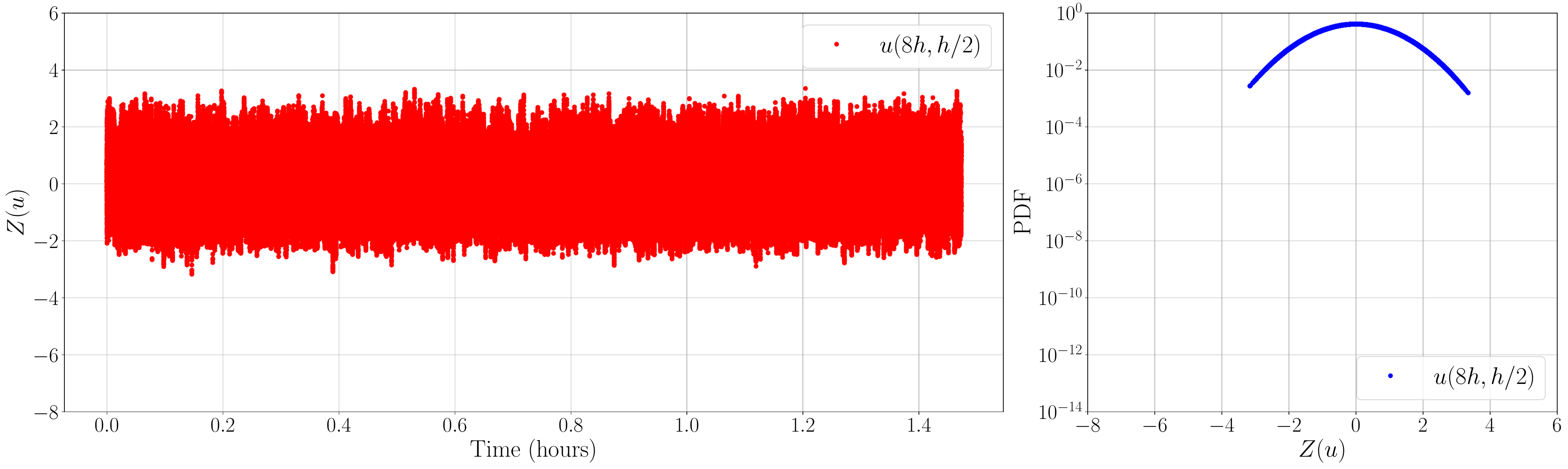}
        \caption{}
        \label{fig:probes_u_3}
    \end{subfigure}    
    \begin{subfigure}{0.48\textwidth}
        \includegraphics[width=\linewidth]{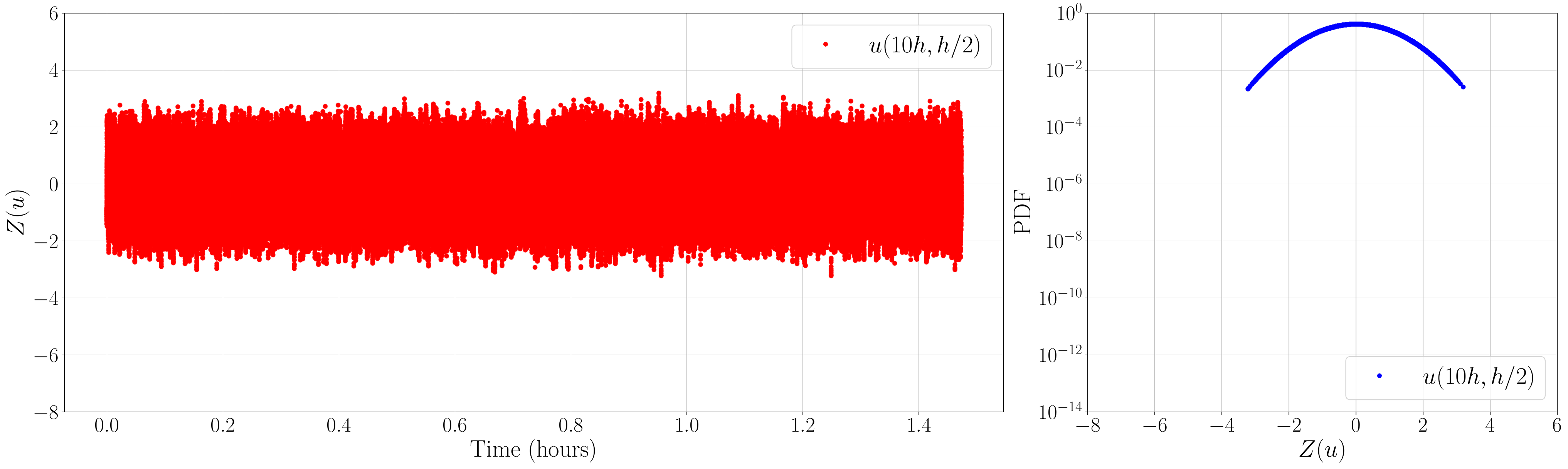}
        \caption{}
        \label{fig:probes_u_4}
    \end{subfigure}

    \caption{Time series and PDF of the streamwise component $u(t)$ of the velocity measured by probes located at $y/h = 0.5$ and in various streamwise positions, ranging from $x/h = 2$ (a), $x/h = 4$ (b), $x/h = 6$ (c), $x/h = 8$ (d), $x/h = 10$ (e). Measurement of 1.5~hours at 100~Hz. }
    \label{fig:probes_u}
\end{figure}

The velocity probes were defined as small square regions of interest (ROIs) of $10\times10~px^2$, which could be placed anywhere in the velocity fields and recorded for as long as needed. At each time step, the velocity components were spatially averaged within the ROI, yielding local probe signals. These probes are particularly useful because they reveal whether intense local events, that may correspond to singular states of the flow, can be found in various regions of the flow. As mentioned previously, these times series can be recorded for hours and then be used as a first preliminary evaluation of the possible existence of extreme events in a region of the flow. In contrast to global quantities (e.g., mean velocity or mean kinetic energy), local probes provide finer interrogation of the flow’s dynamics.

\begin{figure}[h]
\begin{subfigure}{\textwidth}
        \includegraphics[width=\linewidth]{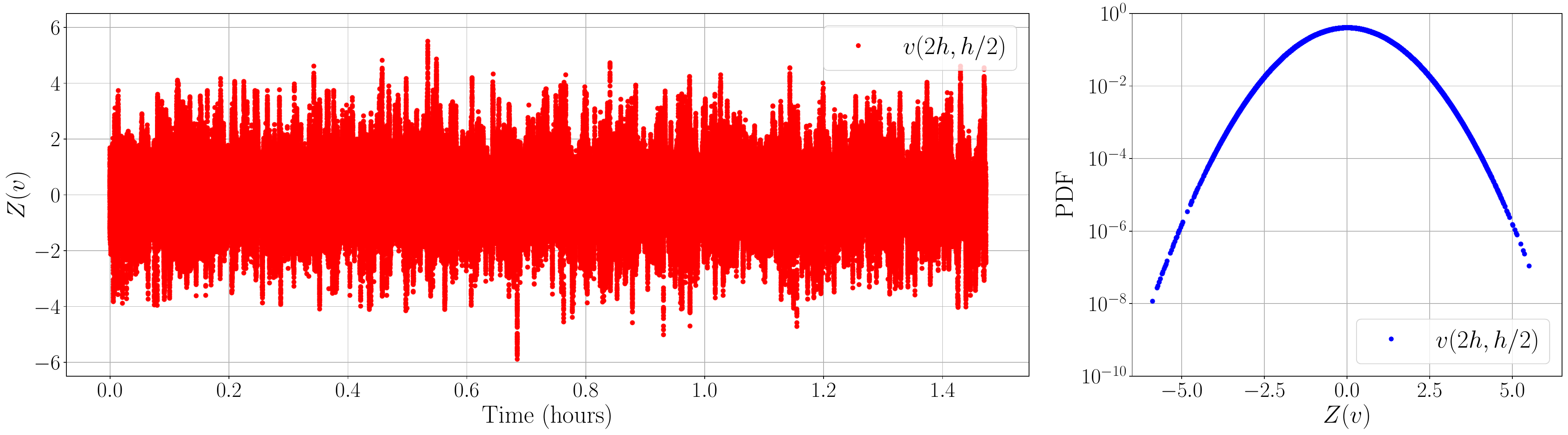} 
        \caption{}
        \label{fig:probes_v_0}
    \end{subfigure}
    \begin{subfigure}{0.48\textwidth}
        \includegraphics[width=\linewidth]{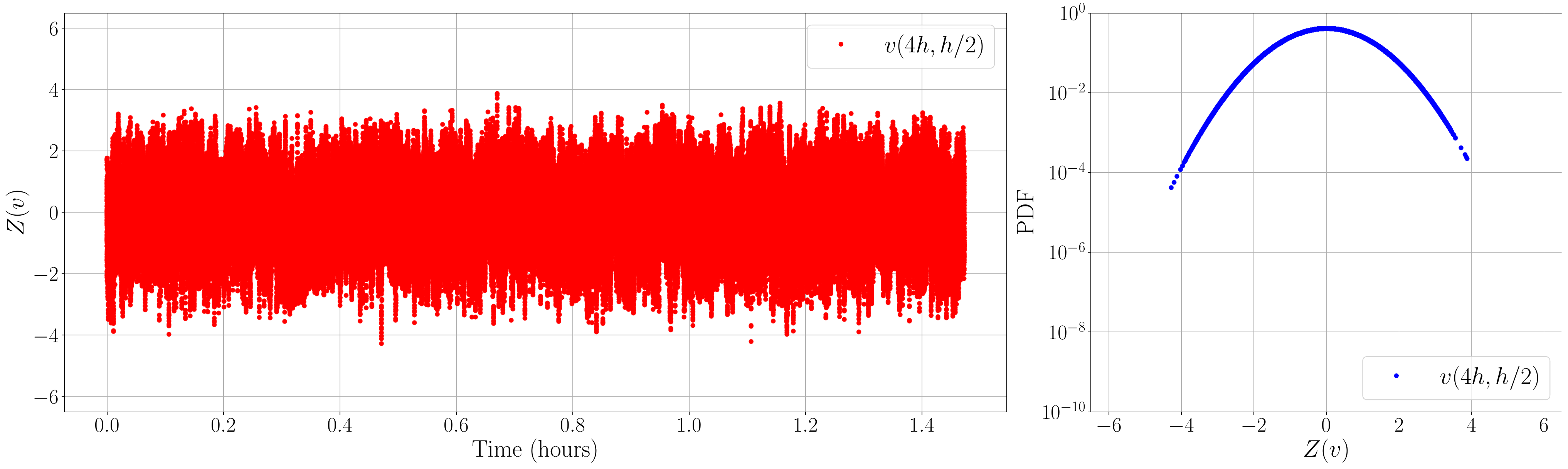}
        \caption{}
        \label{fig:probes_v_1}
    \end{subfigure}   
    \begin{subfigure}{0.48\textwidth}
        \includegraphics[width=\linewidth]{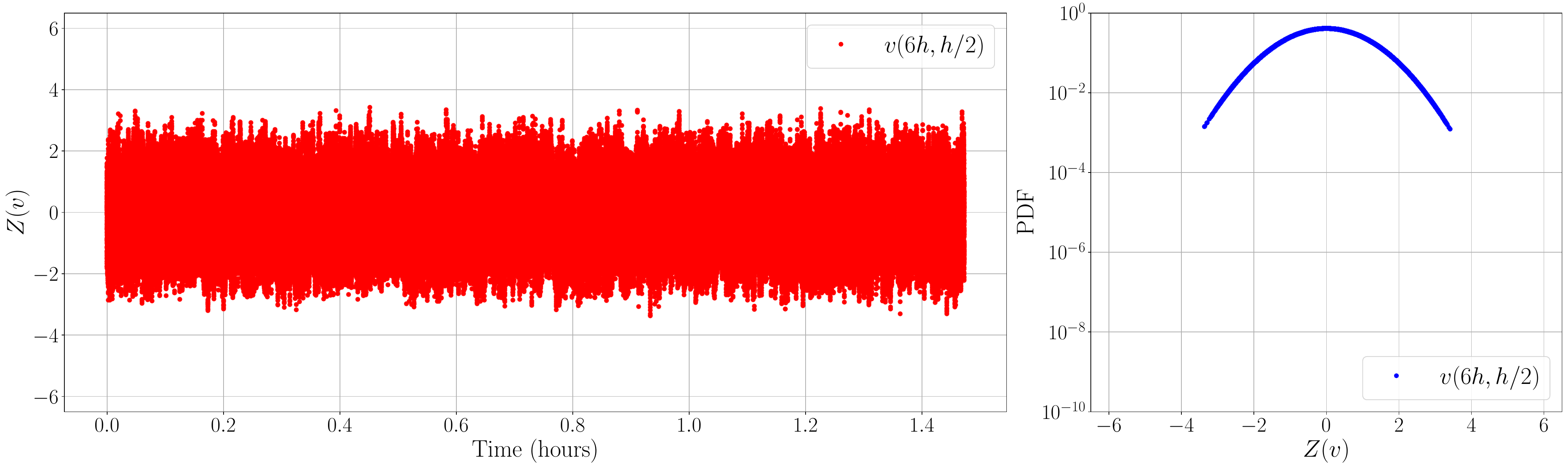}
        \caption{}
        \label{fig:probes_v_2}
    \end{subfigure}      
   \begin{subfigure}{0.48\textwidth}
        \includegraphics[width=\linewidth]{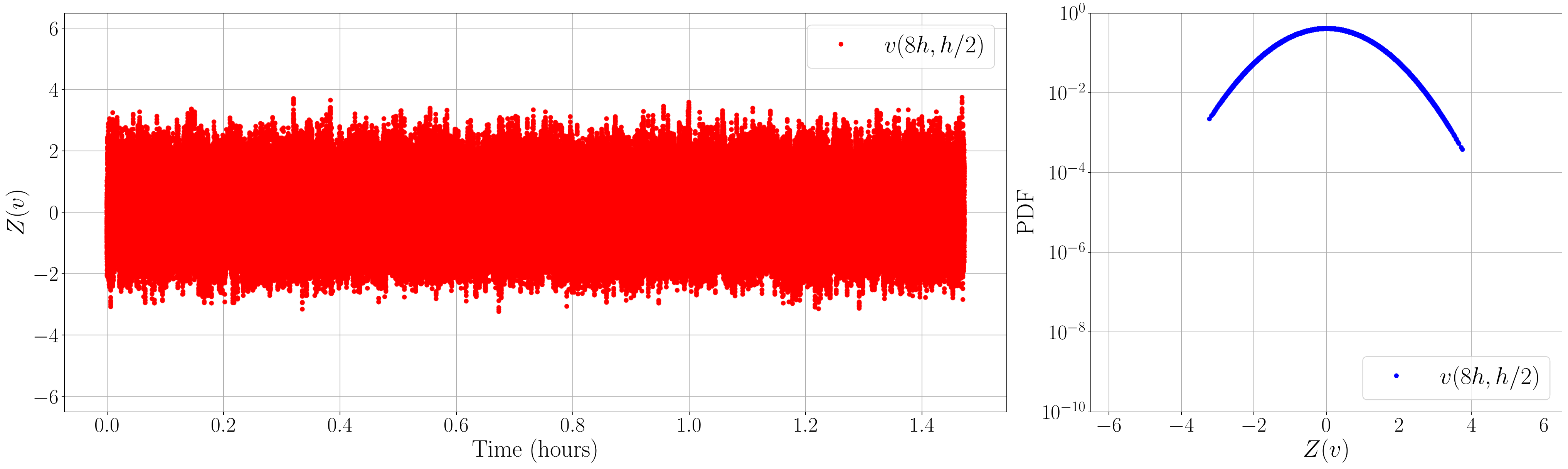}
        \caption{}
        \label{fig:probes_v_3}
    \end{subfigure}    
    \begin{subfigure}{0.48\textwidth}
        \includegraphics[width=\linewidth]{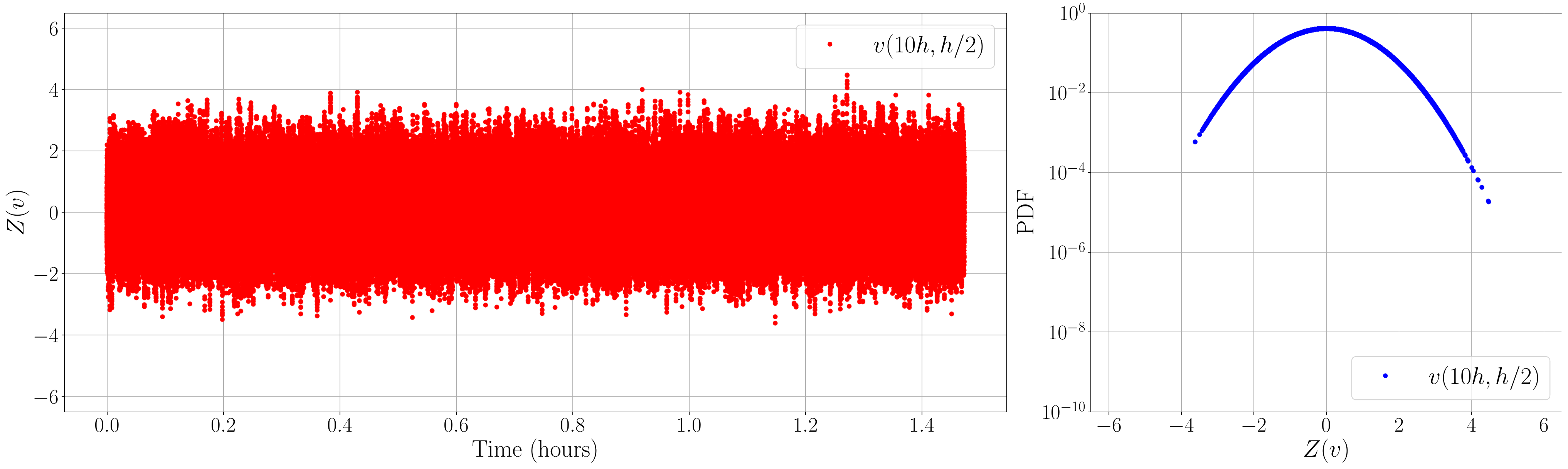}
        \caption{}
        \label{fig:probes_v_4}
    \end{subfigure}  
    
    \caption{Time series and PDF of the wall-normal component $v(t)$ of the velocity measured by probes located at $y/h = 0.5$ and in various streamwise positions, ranging from $x/h = 2$ (a), $x/h = 4$ (b), $x/h = 6$ (c), $x/h = 8$ (d), $x/h = 10$ (e).  Measurements were carried out over 1.5 hours at 100 $Hz$. On the right side of the time-series, their PDF are plotted. }
    \label{fig:probes_v}
\end{figure}

Five velocity probes were defined along an horizontal line at $y/h=0.5$, at streamwise positions ranging from $x/h=2$ to $x/h=10$. \textcolor{black}{They are numbered from 1 ($x/h=2$) to 5 ($x/h=10$).} The reason for the choice of positions of the probes was to detect events potentially related to destabilization of the shear layer, the recirculation region, and the reattachment region downstream of the BFS. Fig.~\ref{fig:VF_probes} shows the locations of the five velocity probes over an instantaneous velocity magnitude field.

Live OFV measurements were then carried out and recorded at $Re_h=2100$ for 1.5 hours at 100 Hz, producing approximately $5.4\times10^5$ samples per probe. For comparison, storing the raw images would have required $\sim 1.7~TB$ of hard drivespace (for images with 8 bits dynamic range). 

The time series were analyzed in terms of the standardized Z-score, defined as  $Z(t) = \frac{X(t) - \mu_t }{\sigma} $, where $\mu_t = \langle X(t) \rangle_t$ represents the time average of time-series X(t)  and $\sigma$ its standard deviation. Extreme events can then be associated to large peak values of $|Z|$ which deviates significantly from the rest of the signal.

\begin{figure}

\begin{subfigure}{\textwidth}
        \includegraphics[width=\linewidth]{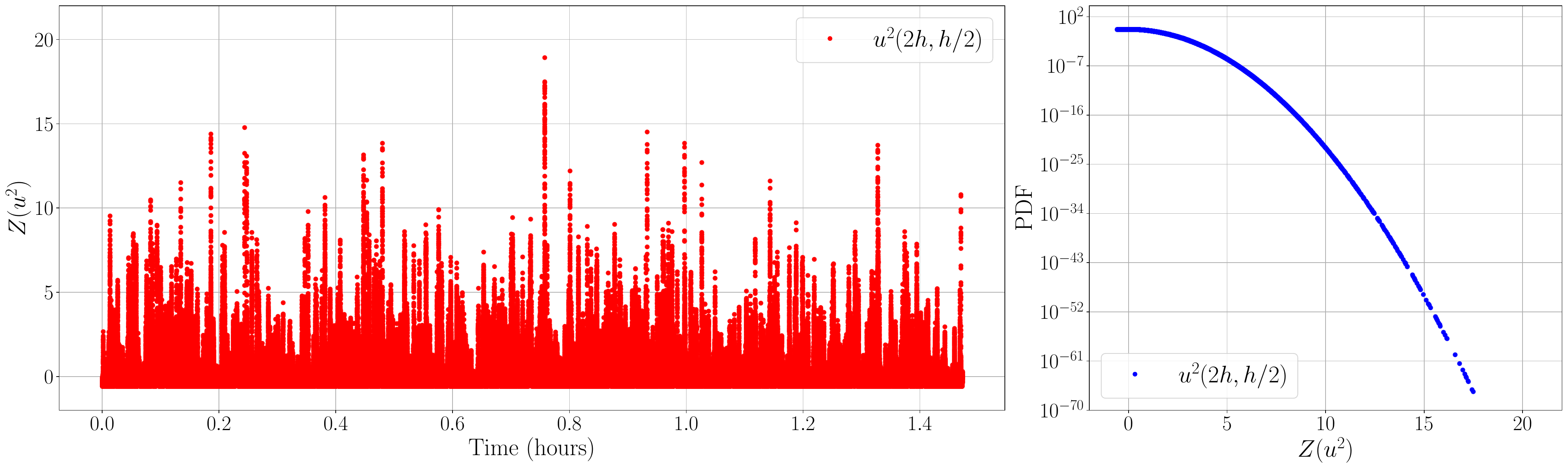} 
        \caption{}
        \label{fig:probes_u2_0}
    \end{subfigure}
    \begin{subfigure}{0.48\textwidth}
        \includegraphics[width=\linewidth]{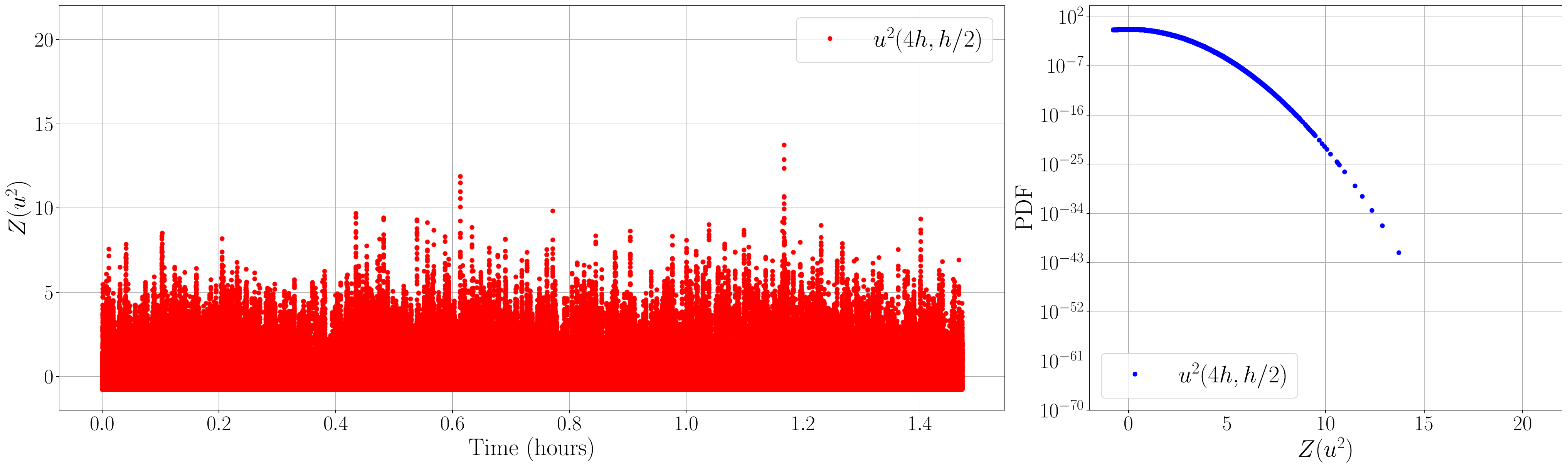}
        \caption{}
        \label{fig:probes_u2_1}
    \end{subfigure}   
    \begin{subfigure}{0.48\textwidth}
        \includegraphics[width=\linewidth]{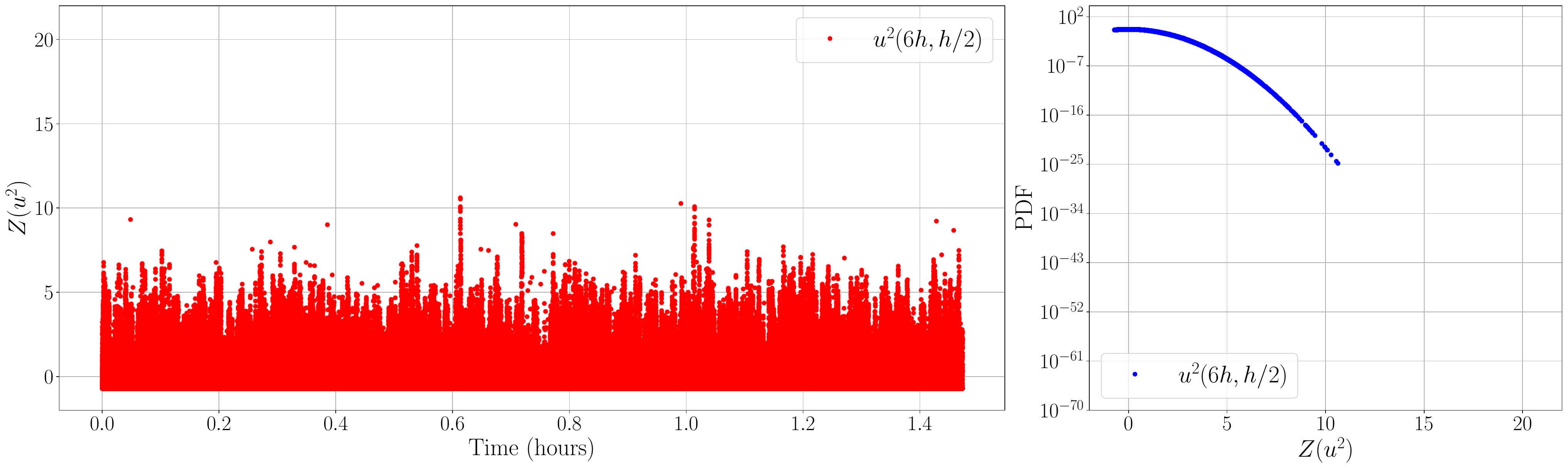}
        \caption{}
        \label{fig:probes_u2_2}
    \end{subfigure}      
   \begin{subfigure}{0.48\textwidth}
        \includegraphics[width=\linewidth]{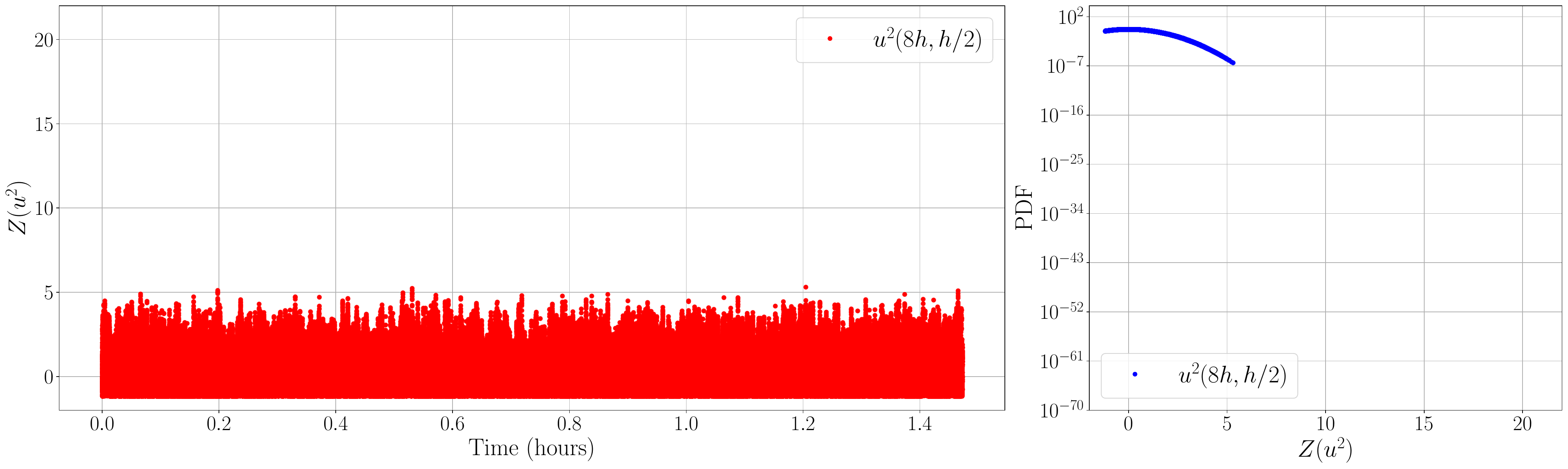}
        \caption{}
        \label{fig:probes_u2_3}
    \end{subfigure}    
    \begin{subfigure}{0.48\textwidth}
        \includegraphics[width=\linewidth]{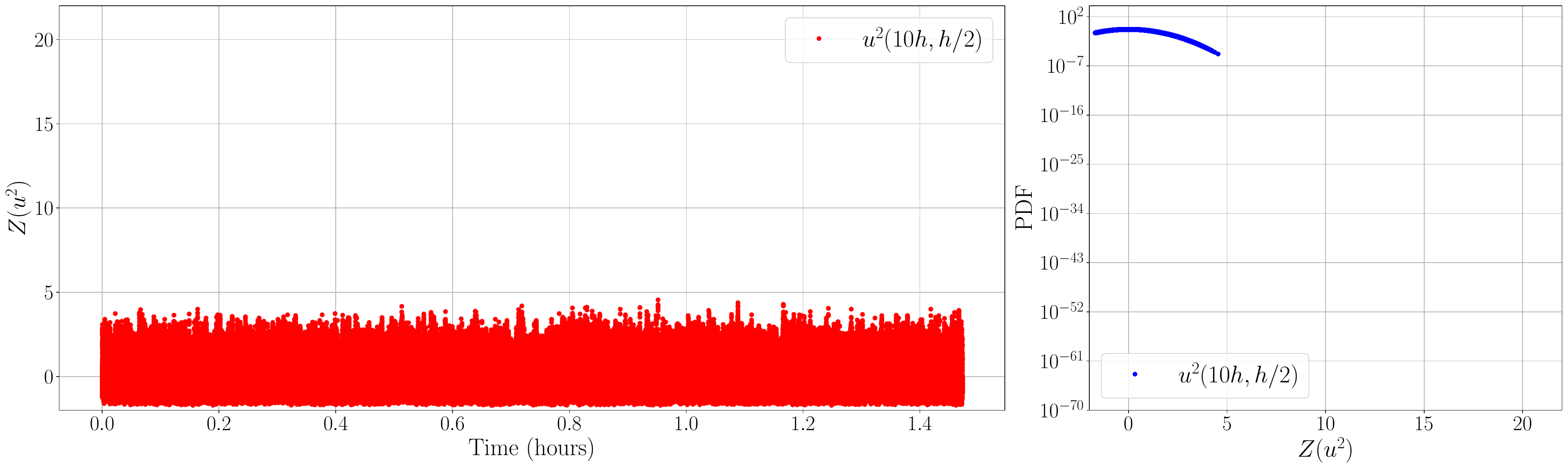}
        \caption{}
        \label{fig:probes_u2_4}
    \end{subfigure} 

    \caption{Time series and PDF of the kinetic energy associated to the streamwise component $u^2(t)$ of the velocity measured by probes located at $y/h = 0.5$ and in various streamwise positions, ranging from $x/h = 2$ (a), $x/h = 4$ (b), $x/h = 6$ (c), $x/h = 8$ (d), $x/h = 10$ (e).  Measurements were carried out over 1.5 hours at 100 $Hz$. On the right side of the time-series, their PDF are plotted.}
    \label{fig:probes_u2}
\end{figure}

Figs.~\ref{fig:probes_u} and ~\ref{fig:probes_v} show respectively the time series and their corresponding probability density functions (PDFs) of the streamwise $(u)$ and wall-normal $(v)$  velocity components measured by the 5 probes. The probe $n^{o}$1 ($(x,y)=(2h,h/2)$) shows large deviations from the mean. The $u$ component (Fig.~\ref{fig:probes_u_0})  exhibits strong peaks of negative streamwise velocity, down to  $Z \approx-6$, confirmed by the long tail of its PDF toward negative values. Similarly,  the $v$ component (Fig.~\ref{fig:probes_v_0})  shows isolated events with $|Z|>5$, including one extreme deviation of $Z\approx-6$.  In  quantitative terms, events with $|Z|>5.5$ correspond to the  99.9903rd percentile for $u$ and 99.9979th percentile for $v$, meaning only $9.6\times10^{-3}\%$ and $2.1\times10^{-3}\%$ of the all samples exceed these thresholds. 

Looking at the time-series of the four other probes located further downstream, no such large deviations can be seen. The fluctuations remain in the range $|Z| <3.5$ and the PDFs are closer to a normal standard symmetric distribution. There is clearly more chances to capture an extreme event with the probe $n^{o}$1 than with the other probes. 

\begin{figure}
\begin{subfigure}{\textwidth}
        \includegraphics[width=\linewidth]{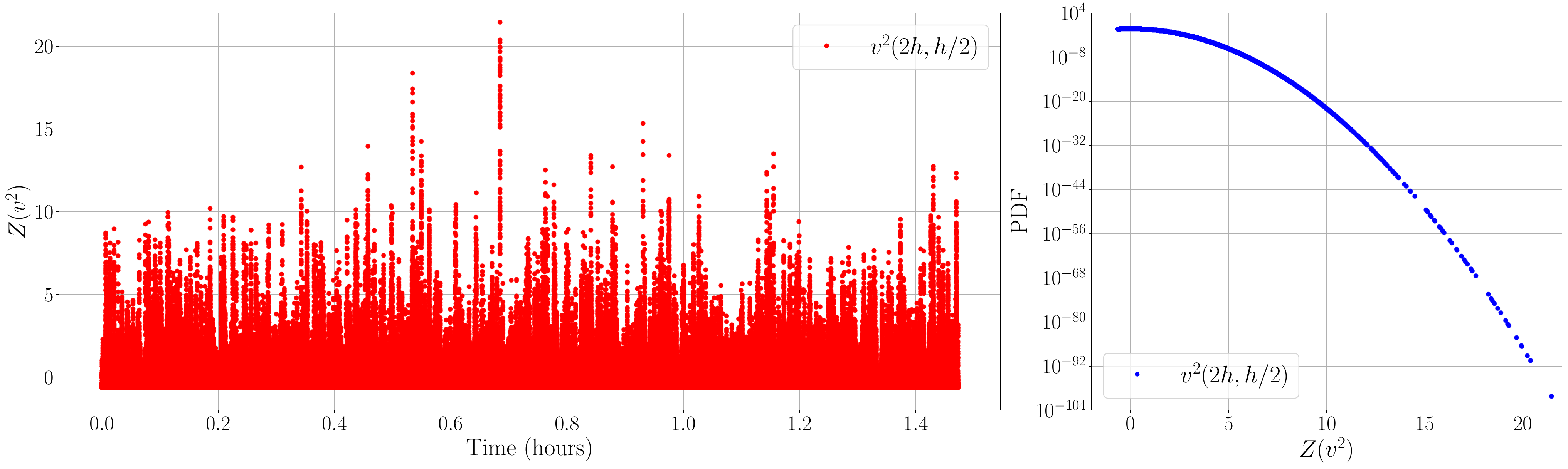} 
        \caption{}
        \label{fig:probes_v2_0}
    \end{subfigure}
    \begin{subfigure}{0.48\textwidth}
        \includegraphics[width=\linewidth]{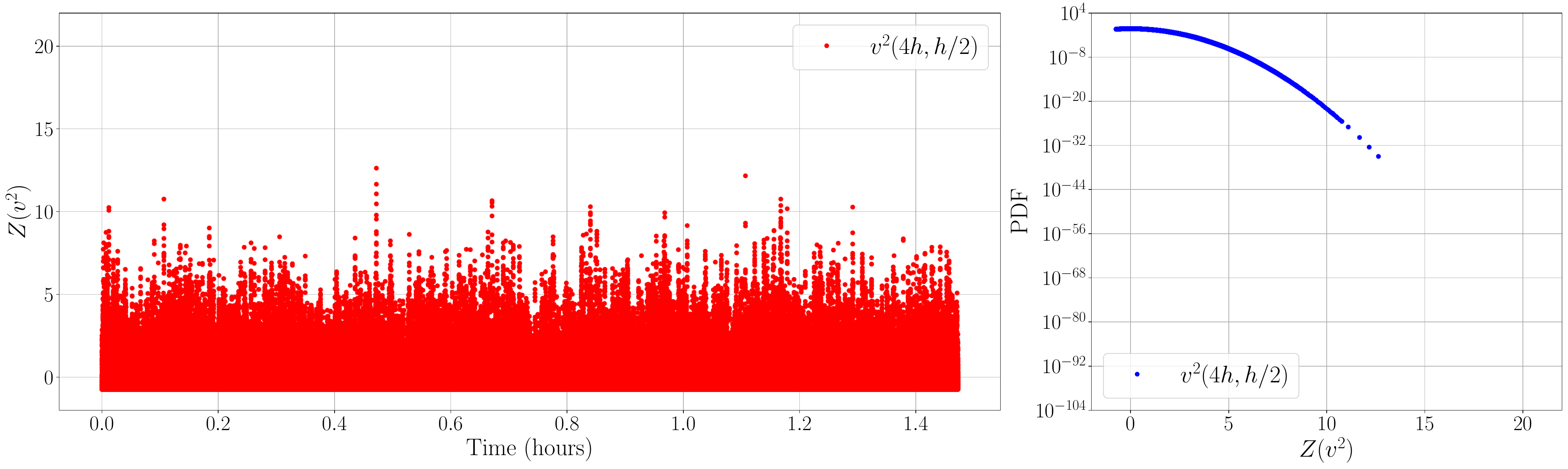}
        \caption{}
        \label{fig:probes_v2_1}
    \end{subfigure}   
    \begin{subfigure}{0.48\textwidth}
        \includegraphics[width=\linewidth]{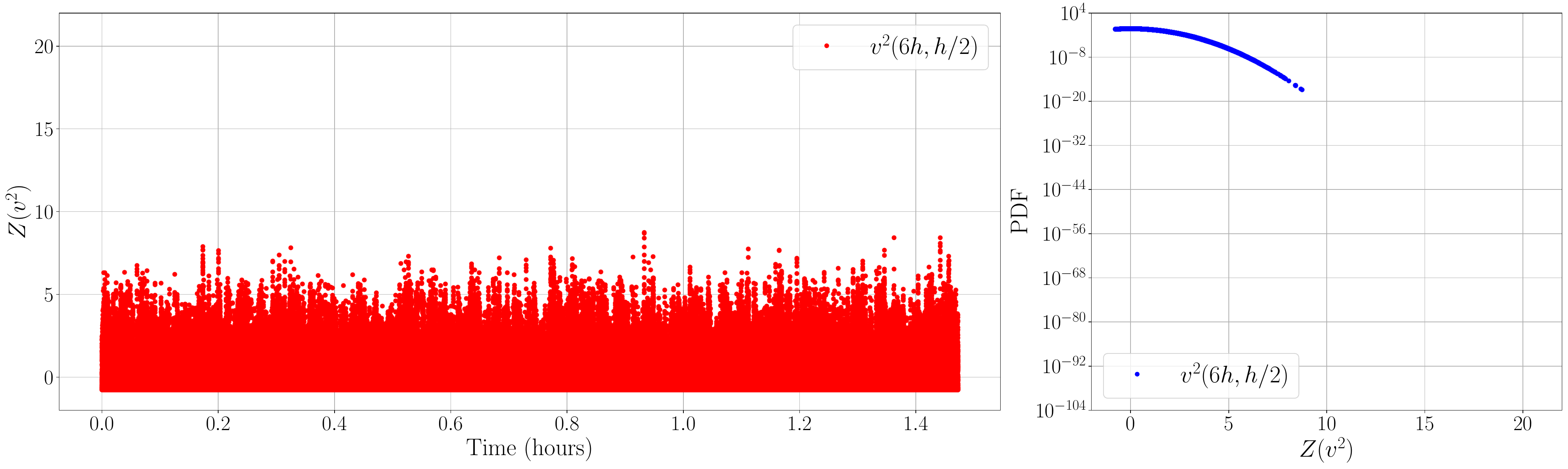}
        \caption{}
        \label{fig:probes_v2_2}
    \end{subfigure}      
   \begin{subfigure}{0.48\textwidth}
        \includegraphics[width=\linewidth]{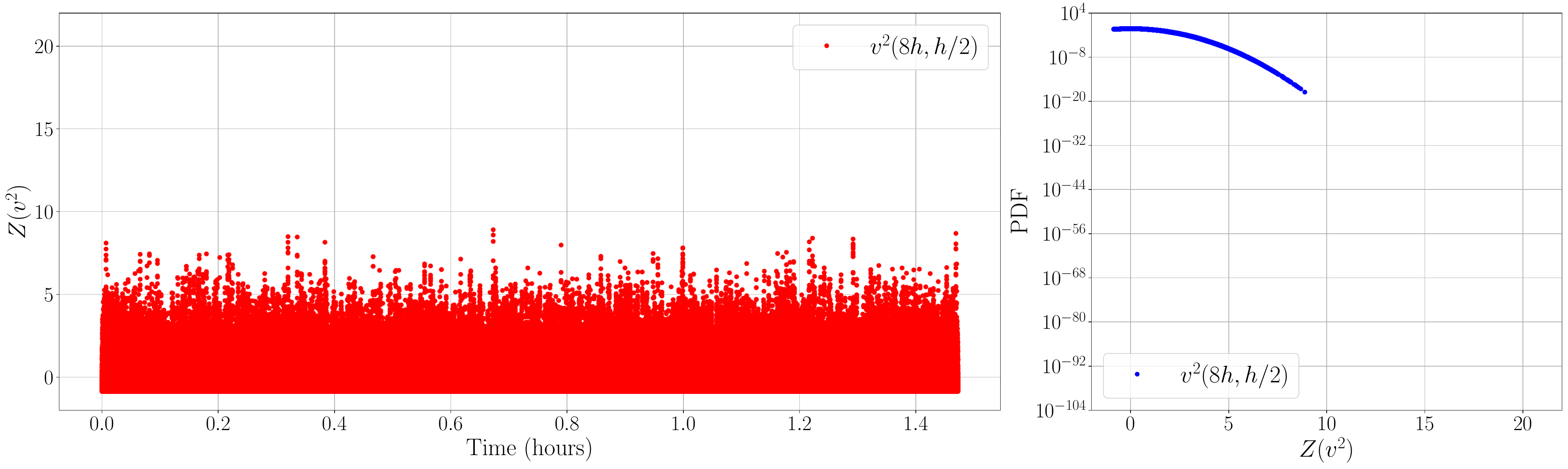}
        \caption{}
        \label{fig:probes_v2_3}
    \end{subfigure}    
    \begin{subfigure}{0.48\textwidth}
        \includegraphics[width=\linewidth]{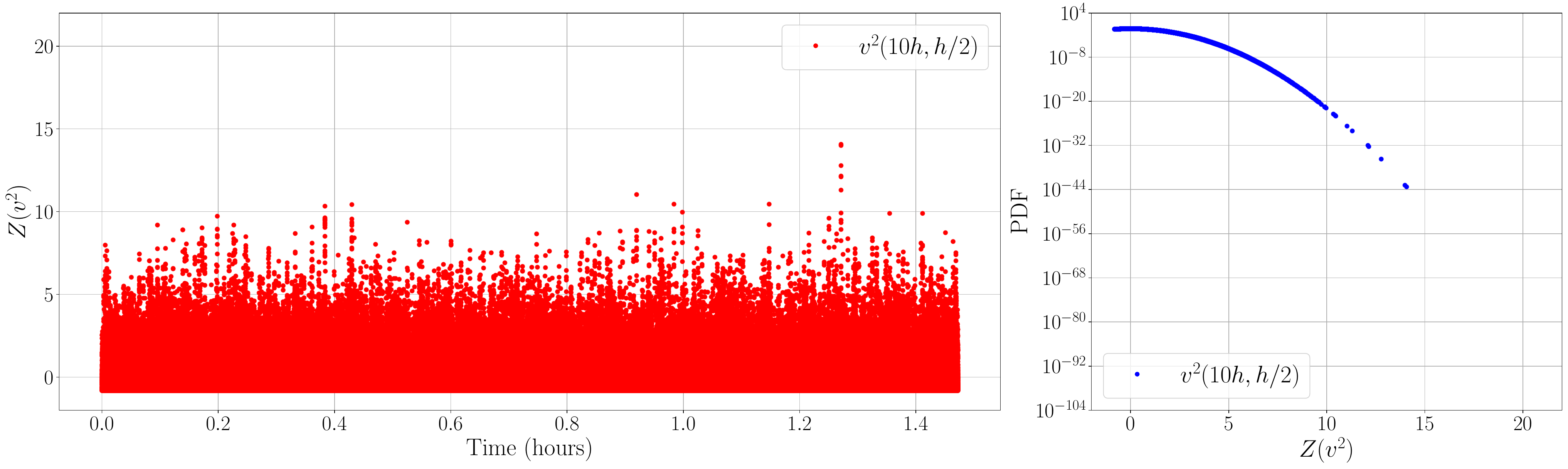}
        \caption{}
        \label{fig:probes_v2_4}
    \end{subfigure}  

    \caption{Time series and PDF of the kinetic energy associated to the wall-normal component $v^2(t)$ of the velocity measured by probes located at $y/h = 0.5$ and in various streamwise positions, ranging from $x/h = 2$ (a), $x/h = 4$ (b), $x/h = 6$ (c), $x/h = 8$ (d), $x/h = 10$ (e).  Measurements were carried out over 1.5 hours at 100 $Hz$.}
    \label{fig:probes_v2}
\end{figure}

Figs.~\ref{fig:probes_u2} and ~\ref{fig:probes_v2} show respectively the corresponding time series and PDFs of the kinetic energy proxies, i.e. $u^2$ and $v^2$. Similarly to the velocity components, strong isolated bursts are found with the probe $n^{o}$1. One can see on Fig.~\ref{fig:probes_u2_0}) one peak of $u^2(t)$ exceeding $Z=15$, while two such events can be seen on Fig.~\ref{fig:probes_v2_0} for $v^2(t)$. These correspond to values above the 99.995th percentile of the samples, confirming their extreme rarity. No such extreme event can be seen in the time-series recorded on the other probes. This is even more evident for the probe $n^{o}$5 (Fig.~\ref{fig:probes_u2_4}), for which the streamwise energy remains within $Z < 5$, with a short-tailed PDF.


The fourth standardized moment (excess kurtosis) was also computed for each probe. Positive excess kurtosis, indicating heavy tails and thus a higher likelihood of extreme events, was found only for the probe $n^{o}$1. For all the other probes excess kurtosis was negative, consistent with platykurtic (flatter-than-Gaussian) distributions and a reduced probability of extreme fluctuations. 

These trends are consistent across all examined statistics: both velocity components and their energy proxies show that extreme deviations are most pronounced for the probe $n^{o}$1.


From this preliminary analysis, we conclude that the most favorable location for detecting rare and extreme events, from those analyzed, is probe $n^{o}$1. Probes at this position were therefore selected as sensors for the live detection protocol described in Sec.~\ref{subsec:detecion_protocol}. \textcolor{black}{The trigger thresholds were then chosen \emph{a priori} based on the tails of the long--duration observation probe distributions} as $Z(u)<-6$ for the streamwise velocity and $Z(v)<-5$ for the wall-normal velocity, \textcolor{black}{so as to target only the deepest excursions observed at that location.}\footnote[1]{The values used for the detection are the raw values of the velocity components that reach the prescribed level of deviations.} \textcolor{black}{These values were used unchanged in the subsequent triggered-acquisition experiment and are introduced here as operational thresholds for event detection, not as universal rare--event boundaries for BFS flows.}

\begin{figure}[H]
    \centering
    \includegraphics[width=\linewidth]{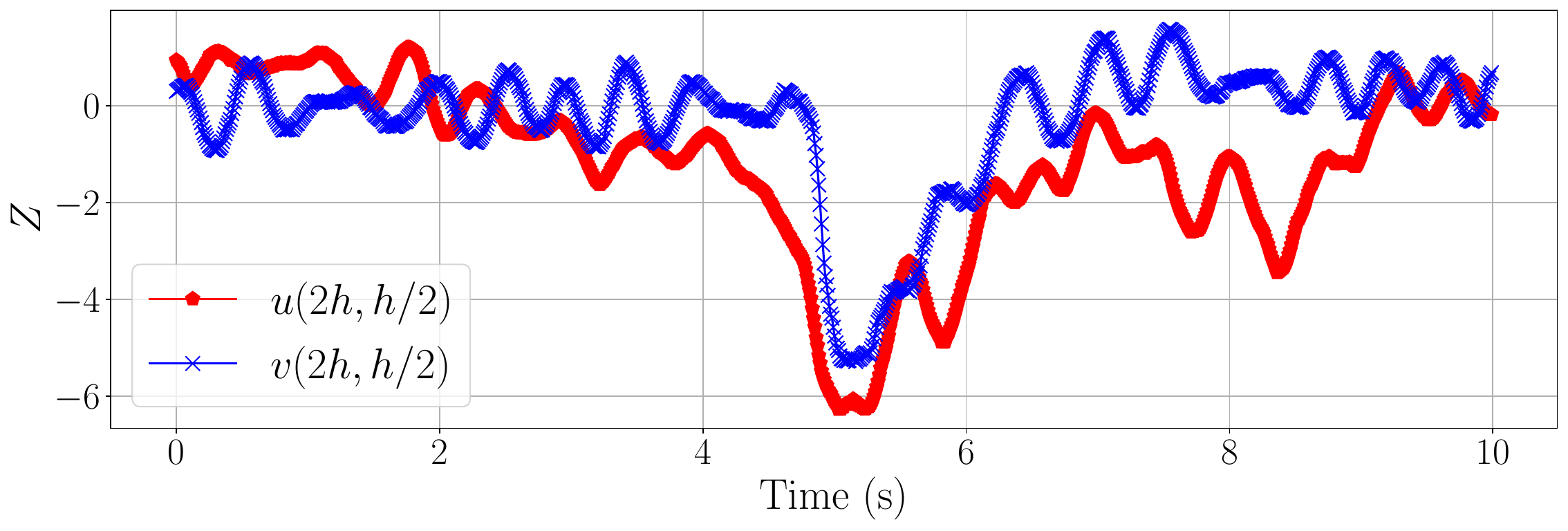}
    \caption{\textcolor{black}{Zoom--in over the velocity probes time series at the the time of larger deviation for each component.}}
    \label{fig:probes_zoom}
\end{figure}

\textcolor{black}{Fig.~\ref{fig:probes_zoom} displays a zoom over the 1000 time steps surrounding the extreme event found in the time-series of both velocity components recorded by the probe $n^{o}$1 (Fig.~\ref{fig:probes_u2}a and Fig.~\ref{fig:probes_v2}a). This figure shows that the large increase of velocity fluctuations develops continuously in time before reaching its peak amplitude. It is therefore related to a real physical evolution of the flow, and is not an isolated processing artifact or a single--sample fluctuation. It is important to note that this extreme event is found in a time-series which was used to define the statistics of velocity fluctuations in different probes. It helped in choosing the thresholds values for the monitoring steps, but no information was stored for further analysis. The extreme event detection and analysis will be detailed in the following section.  }

%% file: Event_detected.tex
\section{Extreme event detection \label{sec:event_detected}}

Using the \textcolor{black}{threshold values found in the preliminary 1.5 hours recording}, a new experiment was \textcolor{black}{then} conducted  to detect an extreme event \textcolor{black}{in real time}. A single event exceeding the thresholds was successfully captured after $1~\mathrm{h}\ 40~\mathrm{min}$ of flow monitoring. To establish a baseline for comparison with the detected event, three additional random acquisitions of 1000 images each were performed. 


\begin{figure}[h]
    \centering
    \begin{subfigure}{\textwidth}
    \centering
        \includegraphics[width=\linewidth]{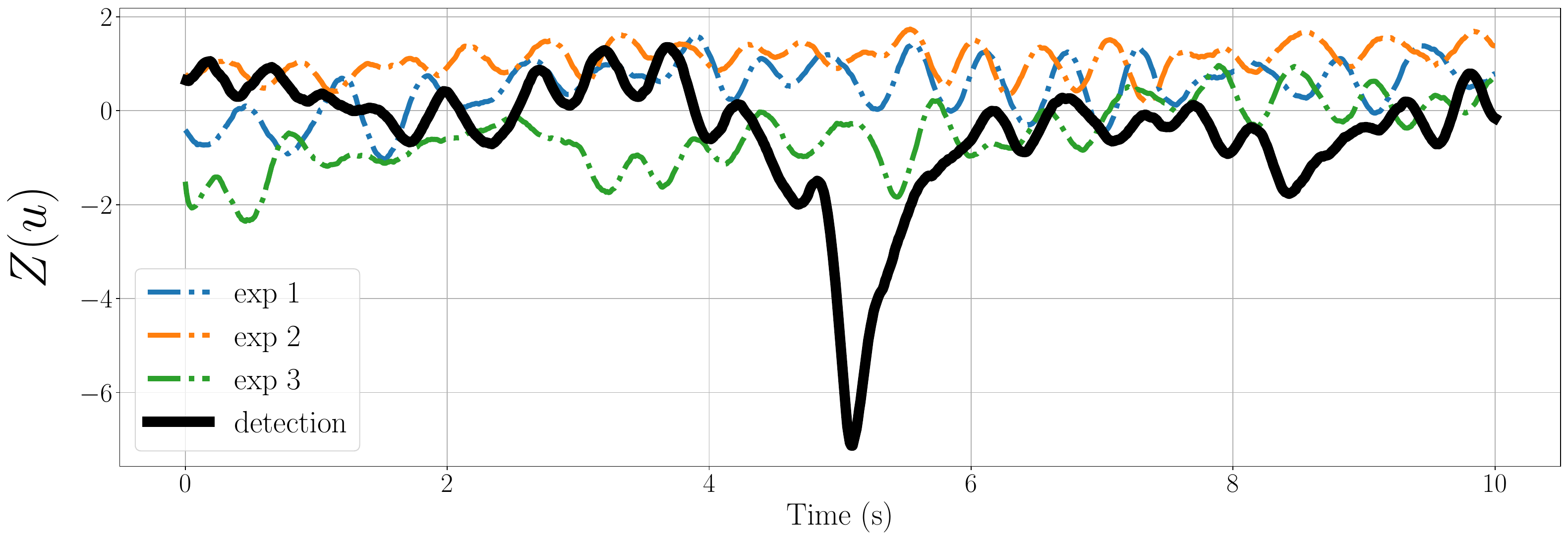}
        \caption{}
        \label{fig:probes_u_comp}    
    \end{subfigure}
    \begin{subfigure}{\textwidth}
        \includegraphics[width=\linewidth]{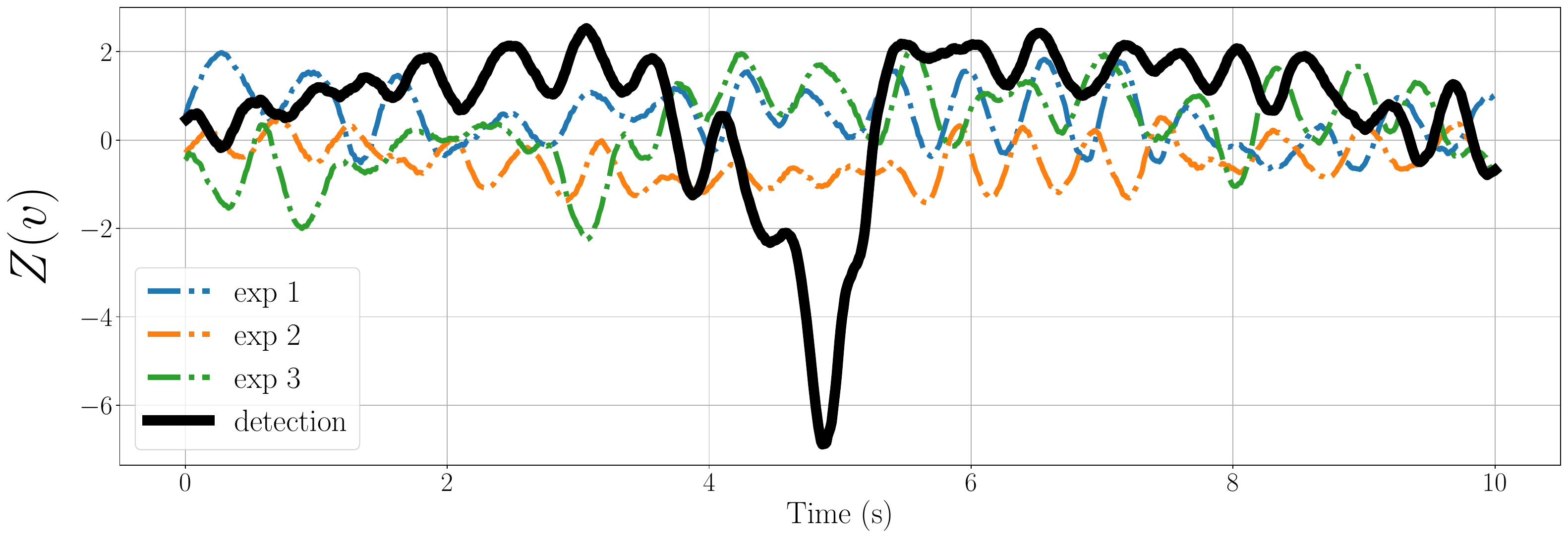}
        \caption{}
        \label{fig:probes_v_comp}    
    \end{subfigure}
    \caption{Comparison of the time series of the velocity components recorded by the probe $n^{o}$1. Exp 1 to 3 are a set of 3 different random experiments used as baselines for comparison with the rare, extreme event detected.}
    \label{fig:probes_comp}
\end{figure}

Fig.~\ref{fig:probes_comp} shows the probe signals corresponding to the detected event (black full line) together with the three random baseline samples (colored dotted lines). The streamwise ($u$) and wall-normal ($v$) components are expressed as Z-scores, as in Sec.~\ref{sec:detection_protocol}. We emphasize that here the mean value $\mu_t$ used to compute $Z$ is obtained using the full recording (1.5~h), which provides a well-converged reference, less susceptible to bias from large fluctuations or limited sampling.

\begin{figure}[h]
    \begin{subfigure}{\linewidth}
        \centering
        \includegraphics[width=\linewidth]{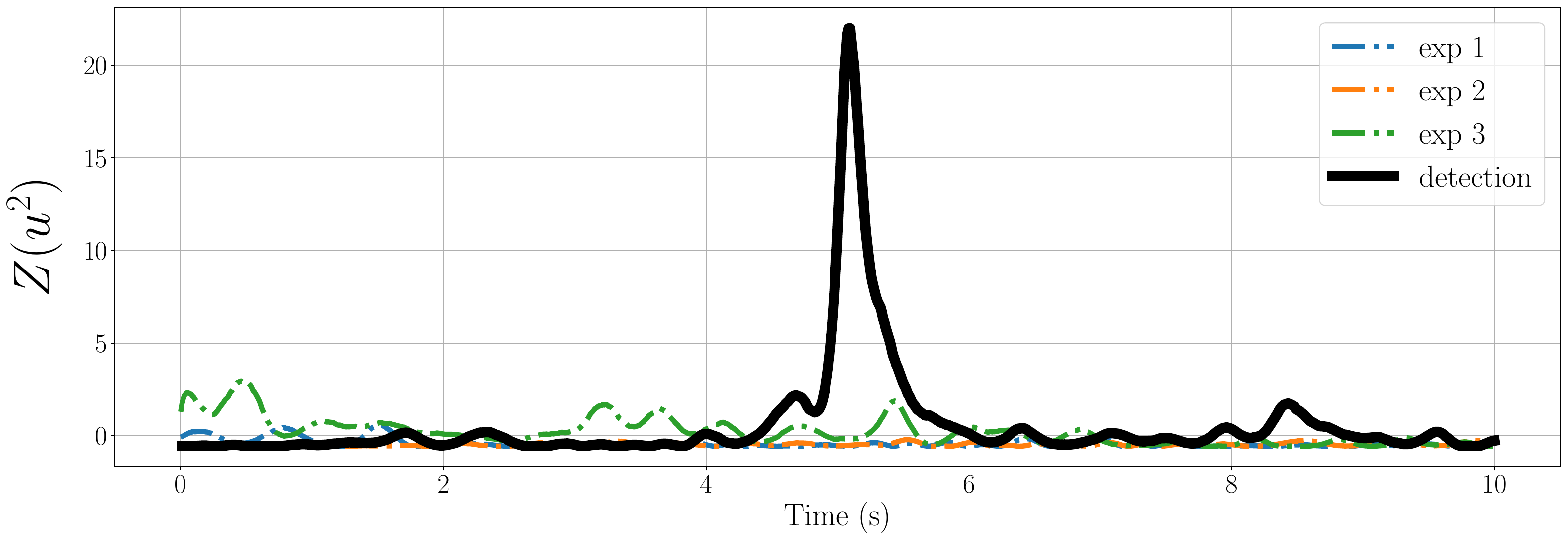}
        \caption{}
        \label{fig:energy_comp_u}    
    \end{subfigure}
    \begin{subfigure}{\linewidth}
        \centering
        \includegraphics[width=\linewidth]{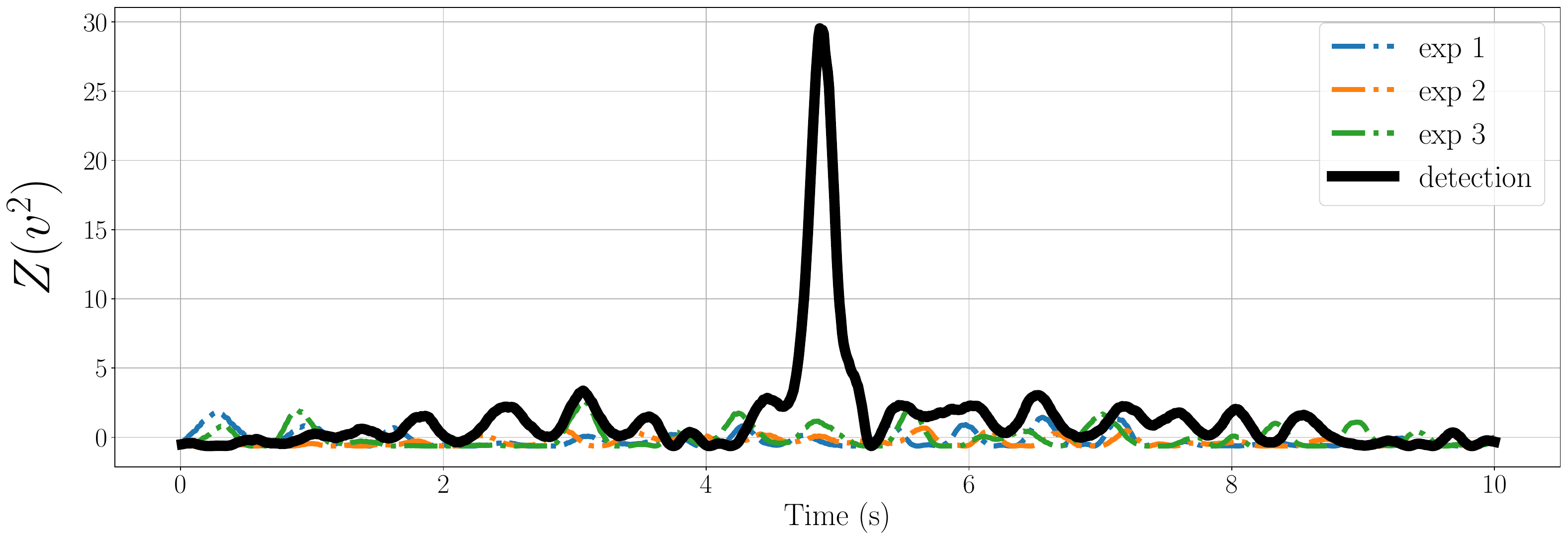}
        \caption{}
        \label{fig:energy_comp_v}    
    \end{subfigure}
    \caption{Comparison of the time-series of the energy associated to each of the velocity components from the probe $n^{o}$1. Exp 1 to 3 are a set of 3 different random experiments used as baselines for comparison with the rare, extreme event detected.}
    \label{fig:energy_comp}
\end{figure}

For the baseline samples (Fig.~\ref{fig:probes_comp}), the local state is oscillatory with small deviations around their mean values in both components. In contrast, the detected event exhibits a large excursion to $Z<-6$ in both $u$ and $v$, indicating a clear departure from the \textit{normal} state of the flow (locally). This departure is even more pronounced in Fig.~\ref{fig:energy_comp}, where the energy proxies reach $Z>20$ for both components, while the baseline acquisitions remain below $Z\approx4$.

\begin{figure}[h]
    \centering
    \includegraphics[width=0.7\linewidth]{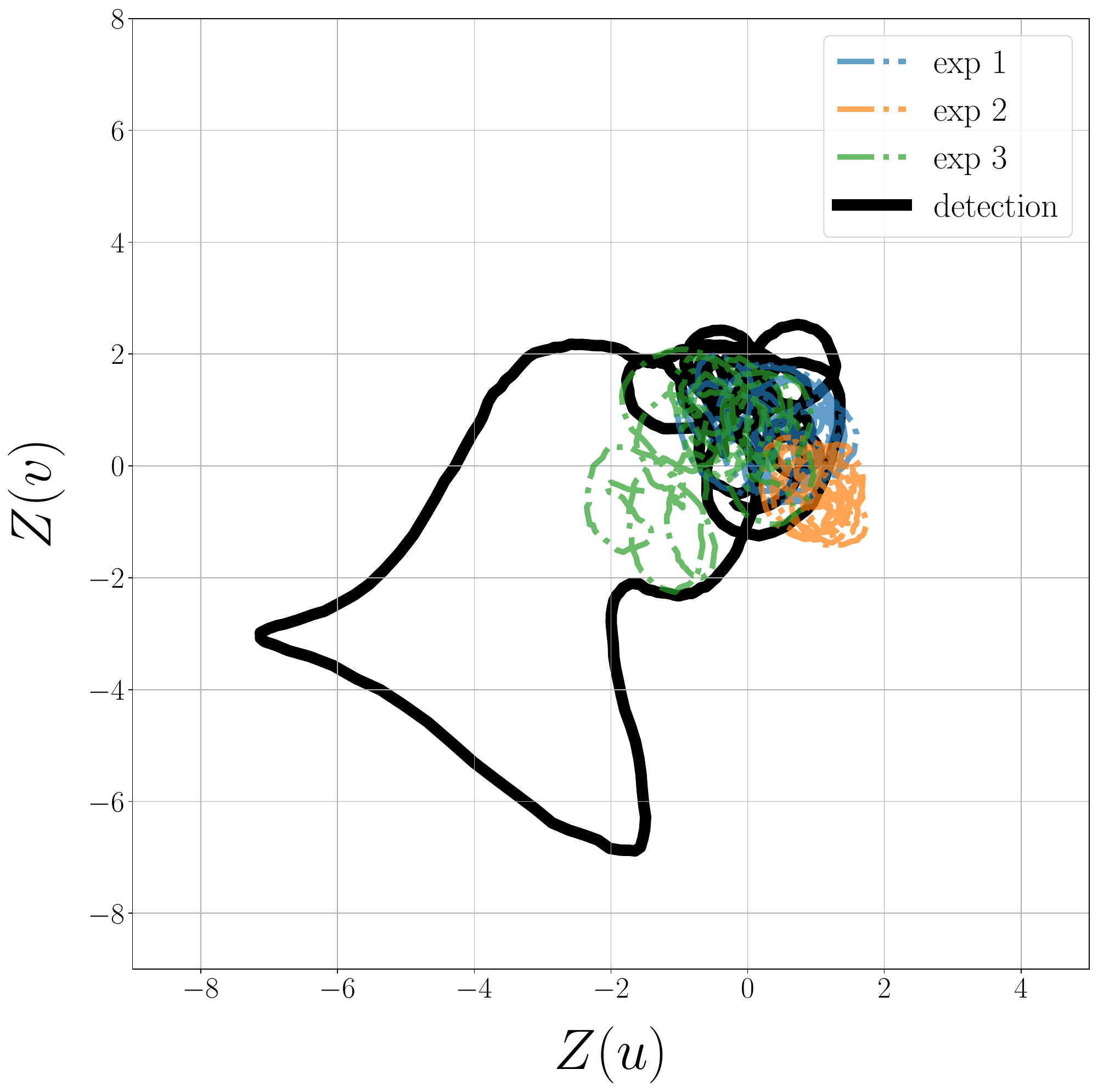}
    \caption{Phase space trajectories $Z_u(Z_v)$ obtained from the velocity probe $n^{o}$1 ($(x,y)=(2h,h/2)$) for the 3 random time series and the extreme event detected. One can see a clear deviation of the phase trajectory of the rare event compared to standard trajectories randomly chosen in the long time series. }
    \label{fig:phase_space}
\end{figure}

Fig.~\ref{fig:phase_space} presents the phase diagram of the velocity time-series measured at probe $n^{o}$1 for the 3 random time-series and the detected event. The random experiments show oscillatory trajectories well centered around a fix point close to the origin. The fix point is not the same for all, however they oscillate in well-defined regions between $Z(u)\approx\pm2$ and $Z(v)\approx\pm2$. In contrast, the detected event exhibits a strange and unique trajectory that departs form its oscillatory behavior around a fixed point, beyond $Z<-6$ in both components. This atypical trajectory is a clear signature of local temporal chaos in dynamical systems \cite{Tailleur2006ProbingRP}, confirming that the fluid flow is taken out of its \textit{normal} state, at least locally. 

The previous results demonstrate that the local state of the flow undergoes a pronounced excursion during the detected event. The question is then: is this strong deviation of the statistics associated to a peculiar event in the flow history ? To try to identify the physical origin of this excursion, the velocity fields and the fluctuating kinetic energy (FKE), recorded around the extreme event, have been analyzed. The FKE is obtained from the fluctuation fields with Reynolds decomposition (Eq.~\ref{eq:re_dec}), computed and normalized by the free-stream kinetic energy (Eq.~\ref{eq:fke}):

\begin{equation}
   {u^{\prime}} = u - \left<u\right>_t 
   \label{eq:re_dec}
\end{equation}
\begin{equation}
    FKE = \left({u^{\prime}}^2 + {v^{\prime}}^2\right)/{U_{\infty}}^2     .
    \label{eq:fke}
\end{equation}

The analysis was limited to the streamwise region $0 \le x/h \le 10$, where the relevant dynamics occur.

\begin{figure}[h!]
    \begin{subfigure}{0.48\linewidth}
        \centering
        \includegraphics[width=\linewidth]{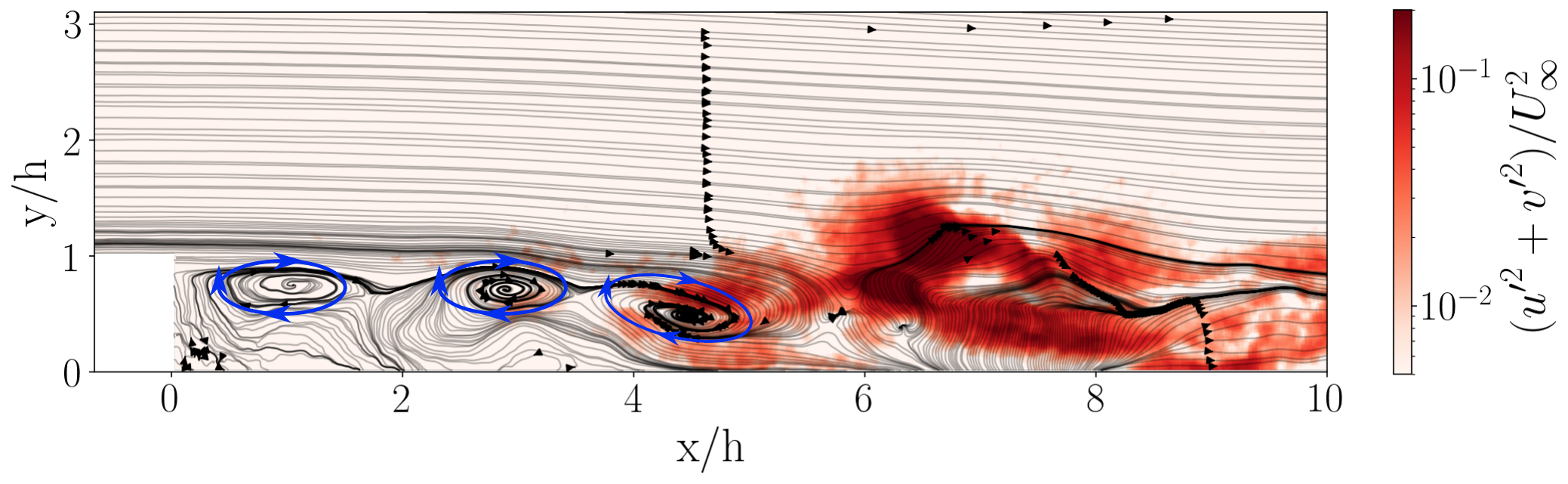}
        \caption{}
        \label{fig:VF_t0}    
    \end{subfigure}
    \begin{subfigure}{0.48\linewidth}
        \centering
        \includegraphics[width=\linewidth]{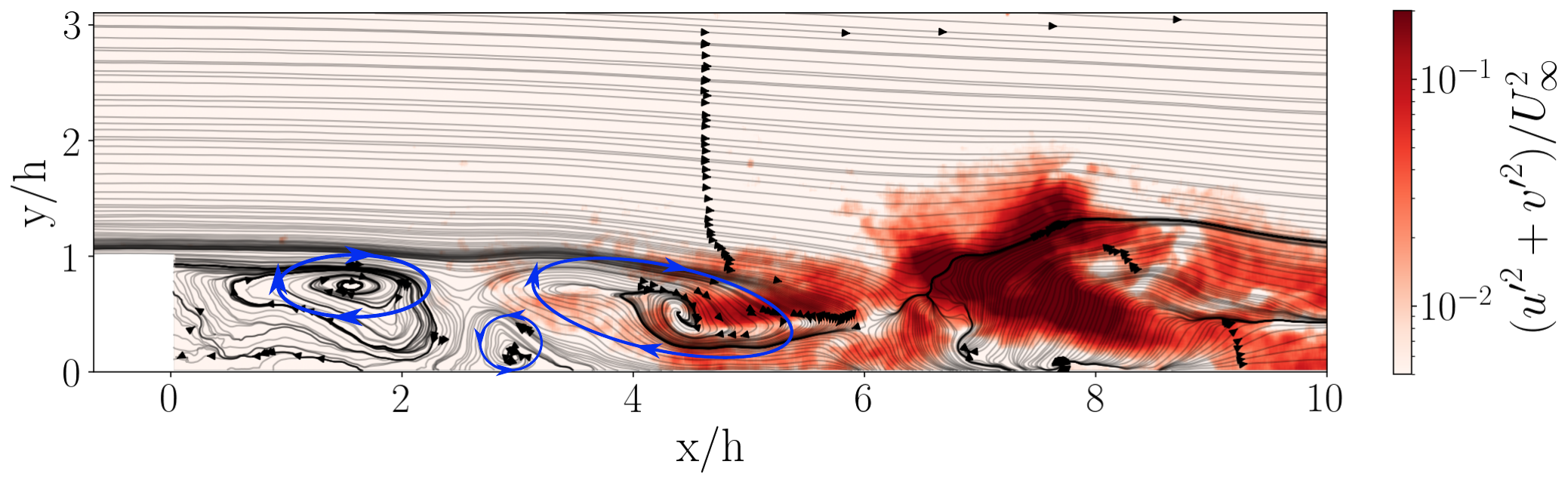}
        \caption{}
        \label{fig:VF_t1}    
    \end{subfigure}

    \begin{subfigure}{0.48\linewidth}
        \centering
        \includegraphics[width=\linewidth]{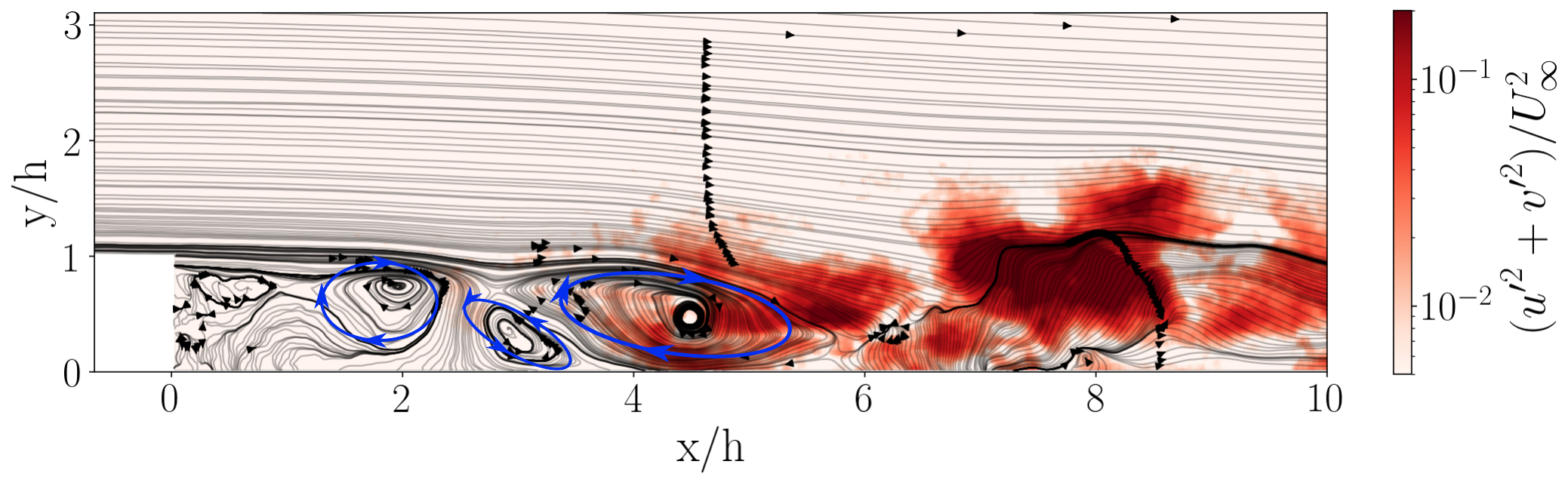}
        \caption{}
        \label{fig:VF_t2}    
    \end{subfigure}
    \begin{subfigure}{0.48\linewidth}
        \centering
        \includegraphics[width=\linewidth]{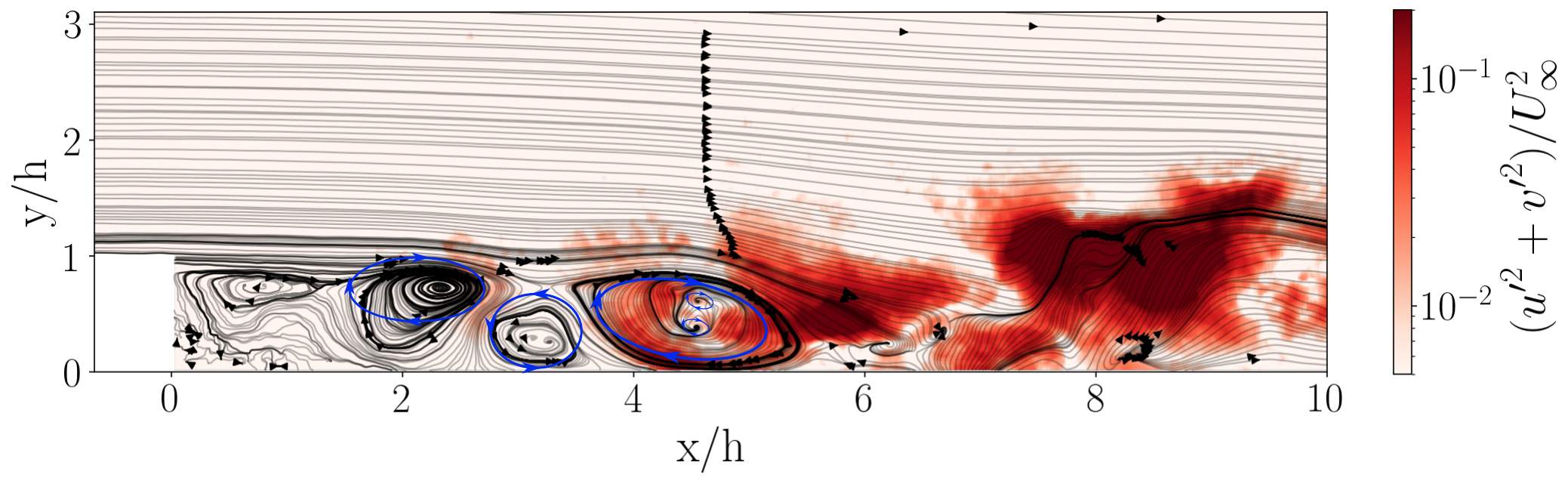}
        \caption{}
        \label{fig:VF_t3}    
    \end{subfigure}
    \begin{subfigure}{0.48\linewidth}
        \centering
        \includegraphics[width=\linewidth]{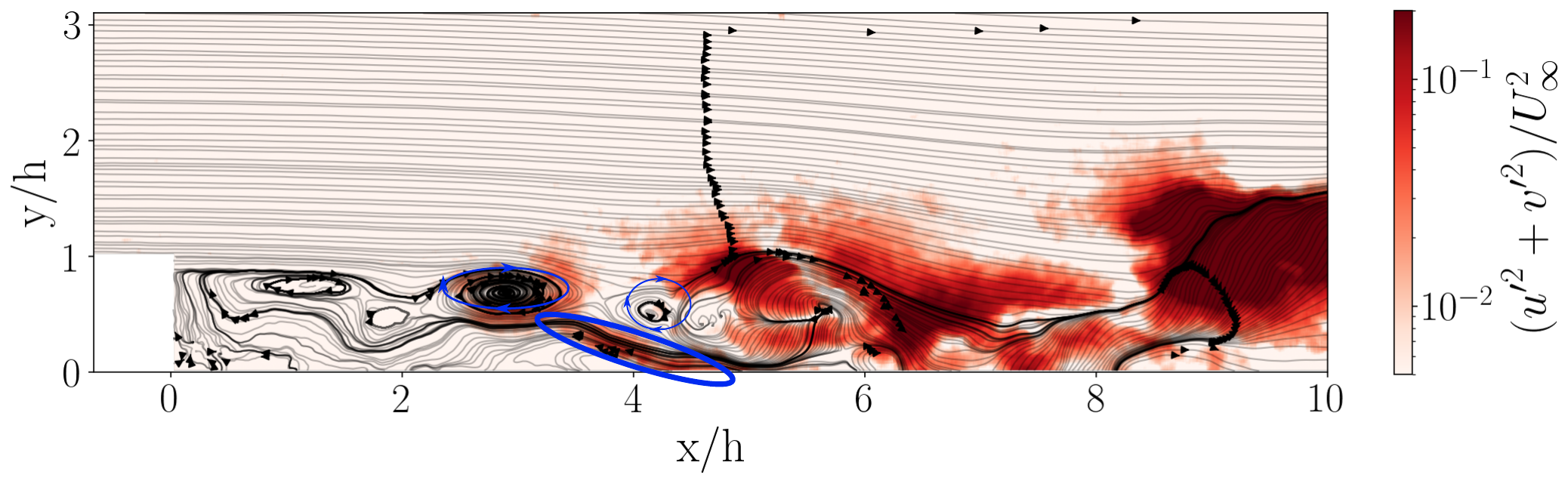}
        \caption{}
        \label{fig:VF_t4}    
    \end{subfigure}
    \begin{subfigure}{0.48\linewidth}
        \centering
        \includegraphics[width=\linewidth]{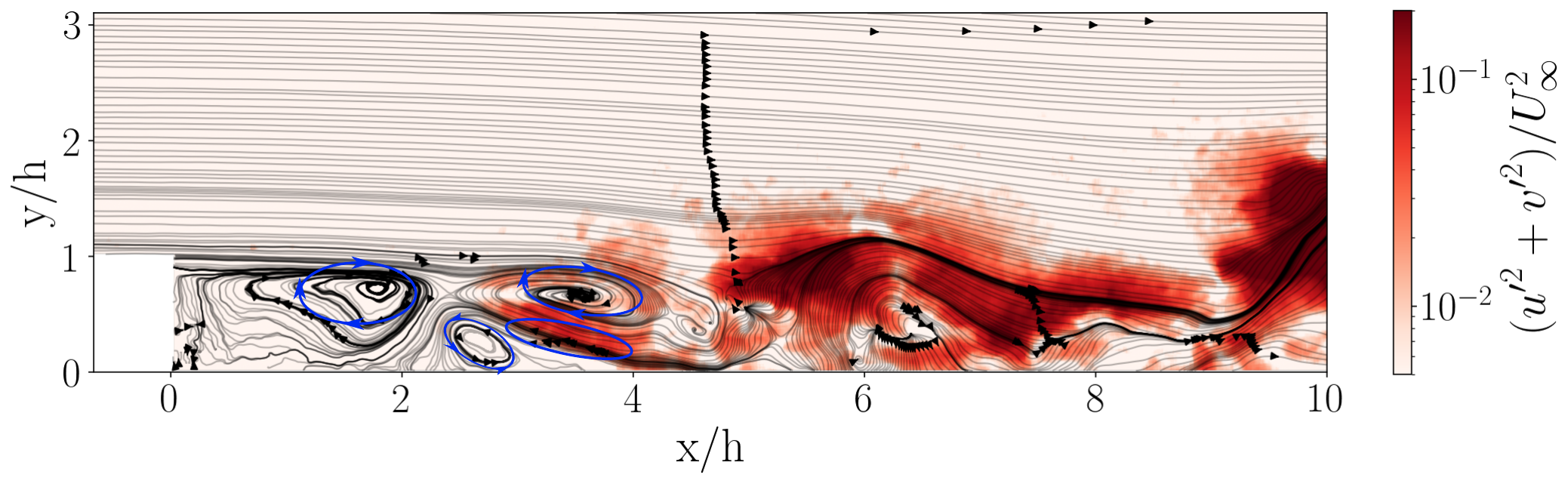}
        \caption{}
        \label{fig:VF_t5}    
    \end{subfigure}
    \begin{subfigure}{0.48\linewidth}
        \centering
        \includegraphics[width=\linewidth]{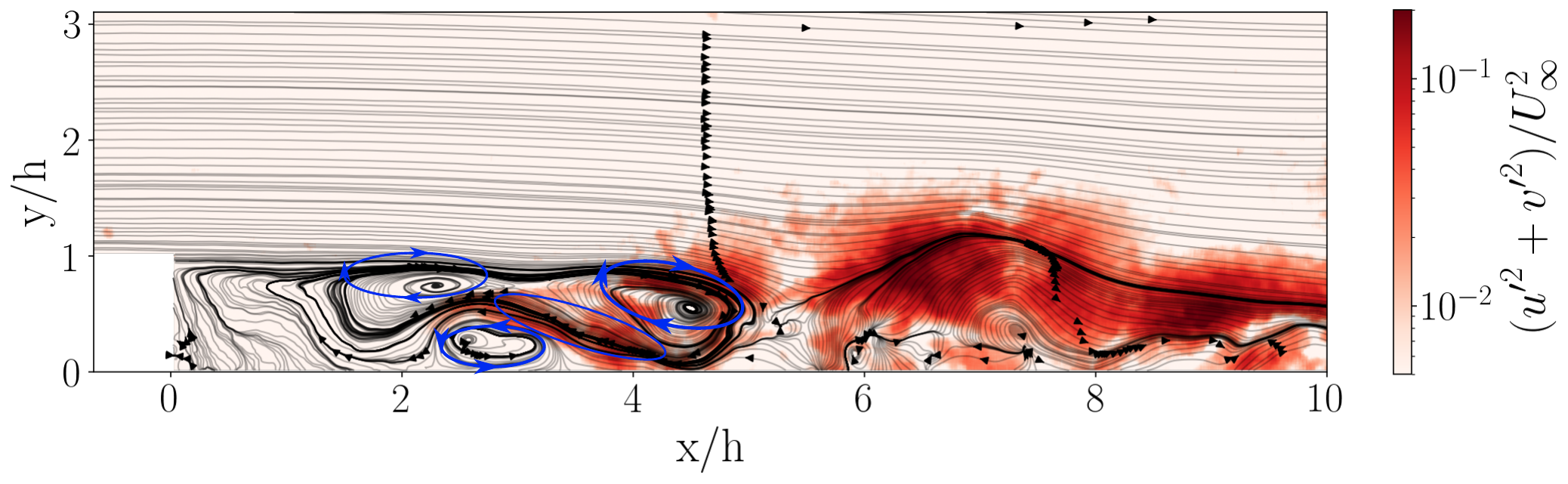}
        \caption{}
        \label{fig:VF_t6}    
    \end{subfigure}
    \begin{subfigure}{0.48\linewidth}
        \centering
        \includegraphics[width=\linewidth]{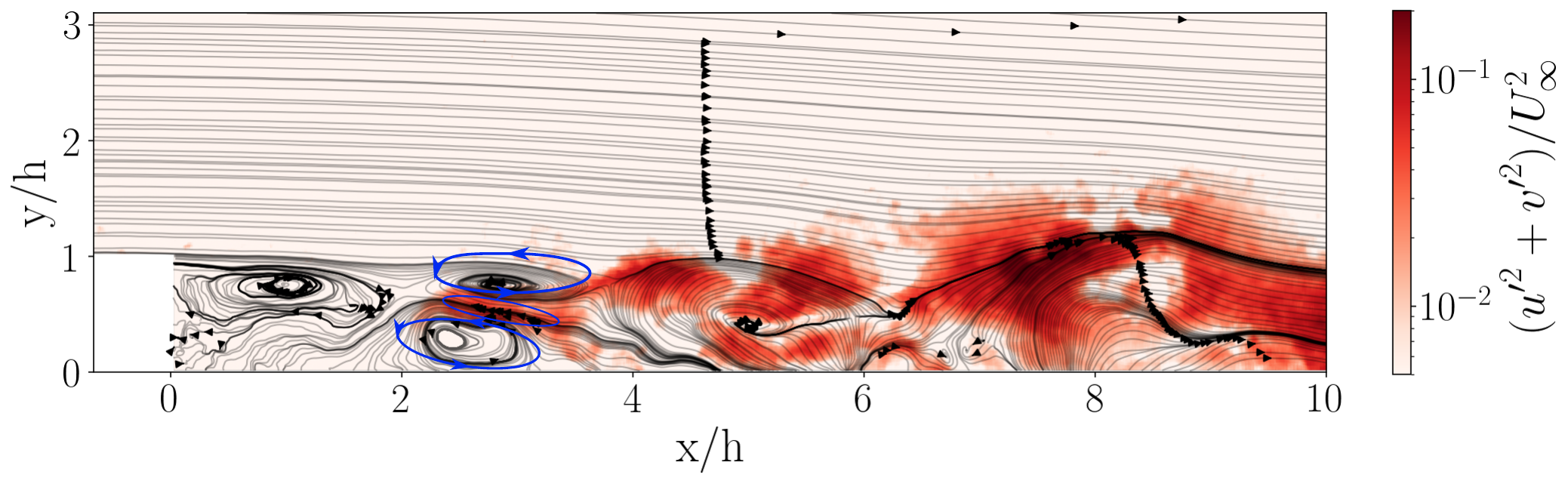}
        \caption{}
        \label{fig:VF_t7}    
    \end{subfigure}
    \begin{subfigure}{0.48\linewidth}
        \centering
        \includegraphics[width=\linewidth]{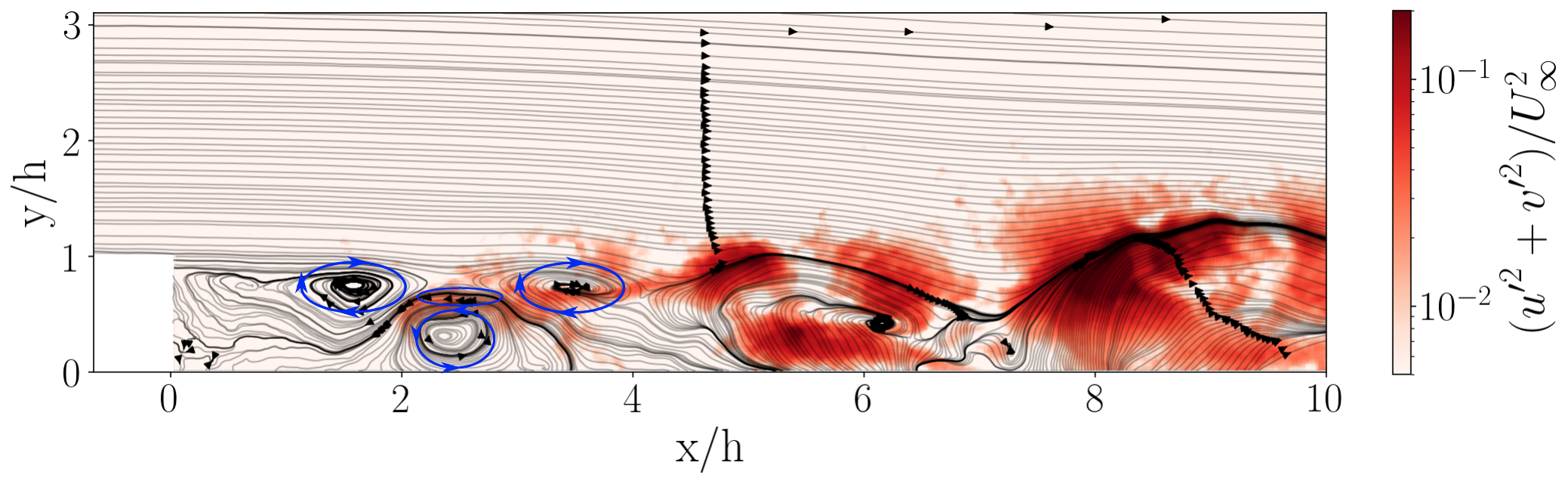}
        \caption{}
        \label{fig:VF_t8}    
    \end{subfigure}
    \begin{subfigure}{0.48\linewidth}
        \centering
        \includegraphics[width=\linewidth]{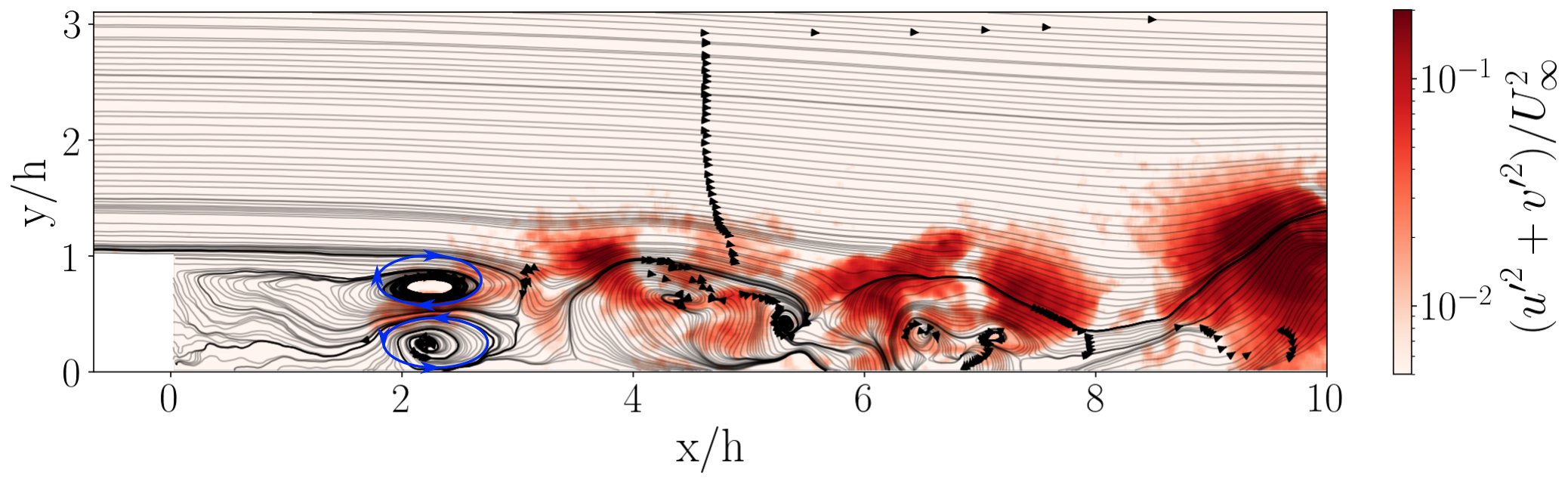}
        \caption{}
        \label{fig:VF_t9}    
    \end{subfigure}
    \caption{Fluctuating kinetic energy snapshots and streamline plots. The snapshots are extracted from times $t=3.65,3.80,3.90,4.00,4.20,4.35,4.55,4.75,4.90,5.15~s$ from the image series containing the rare/extreme event, respectively with the order of the images. Event was detected at $t=5~s$. \textcolor{black}{Movie online.}}
    \label{fig:fke}

\end{figure}

Figure~\ref{fig:fke} shows ten successive FKE instantaneous contour plots superposed to the streamlines of the flow, over the time interval ranging from  $t=3.65$ to $t=5.15~\mathrm{s}$, centered around the detected event. The analysis of the time evolution of the flow indicates that the event can be described as an upstream-directed horizontal jet that penetrates the recirculation region, propelled by counter-rotating vortices located between the lower wall and the shear layer over $x/h\approx5$–$3$. The jet ultimately rolls up into a slowly rotating vortex that persists near $x/h\approx2$.

A more detailed chronology is as follows. In Fig.~\ref{fig:VF_t0}, three Kelvin–Helmholtz (KH) vortices are visible along the shear layer at $x/h\approx(1,\,3,\,4.5)$; FKE is weak near the two downstream structures and larger beyond $x/h=6$. In Fig.~\ref{fig:VF_t1}, the two downstream KH vortices merge into a larger structure, while a small vortex emerges near the lower wall at $x/h\approx3$; the FKE grows around the merged vortex. In Fig.~\ref{fig:VF_t2}, the merged vortex intensifies and the near-wall vortex expands. In Fig.~\ref{fig:VF_t3}, the merged vortex remains approximately in place (over $\sim0.25~\mathrm{s}$), with elevated FKE and a transient double core; simultaneously, the near-wall vortex continues to grow and the most upstream KH structure (now at $x/h\approx2.5$) strengthens. In Fig.~\ref{fig:VF_t4}, the merged vortex collapses, producing strong FKE and initiating a backward (upstream) injection of flow into the recirculation region (highlighted by the blue ellipse). In Fig.~\ref{fig:VF_t5}, this injection advances upstream, with the local FKE maximum shifting from $x/h\approx4$ to $x/h\approx3.5$; it interacts with the shear-layer vortex and begins to couple with the near-wall vortex at $x/h\approx2.8$. In Fig.~\ref{fig:VF_t6}, the injection continues upstream, propelled by interactions between the near-wall vortex and successive KH vortices shed from the shear layer; the FKE field shows the fluctuation energy moving with the injected mass inside the recirculation zone. In Fig.~\ref{fig:VF_t7}, the counter-rotation of the KH and near-wall vortices enhances the upstream propulsion. In Fig.~\ref{fig:VF_t8}, the injection persists, fueled by the coupled dynamics of the upper and lower vortices, which are advected in different directions and at different velocities; by this time, the passing structure has already produced the large negative excursion in the local $v$ signal ($Z<-6$). Finally, in Fig.~\ref{fig:VF_t9}, the injected mass appears to merge with the near-wall vortex, which grows and persists longer than the other structures. 
The full dynamics of the flow can be observed in the video in the supplementary material.

\begin{figure}
    \centering
    \begin{subfigure}{0.49\linewidth}
        \includegraphics[width=\linewidth]{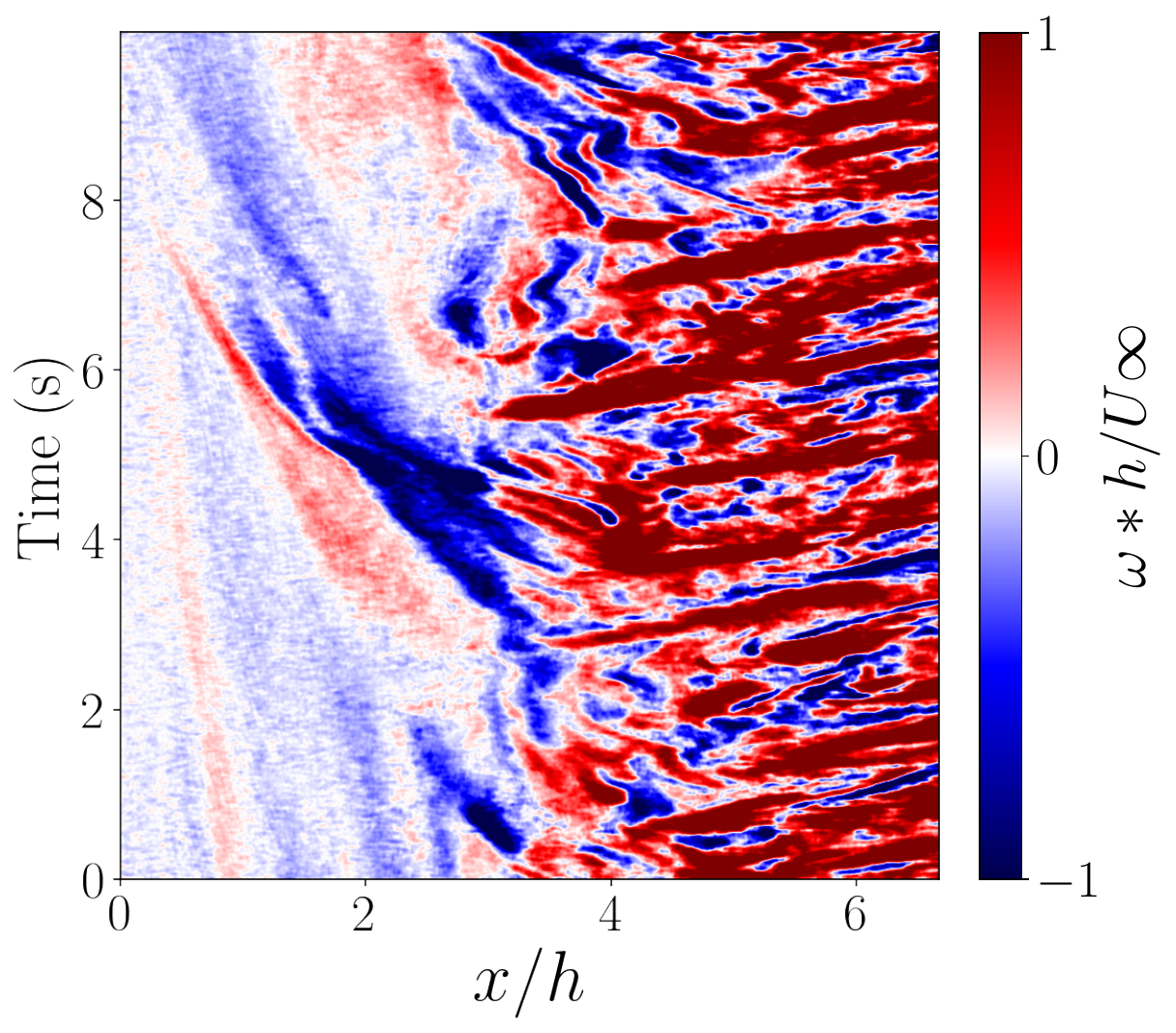}
        \caption{}
        \label{fig:space_time_vort}    
    \end{subfigure}
    \begin{subfigure}{0.49\linewidth}
        \includegraphics[width=\linewidth]{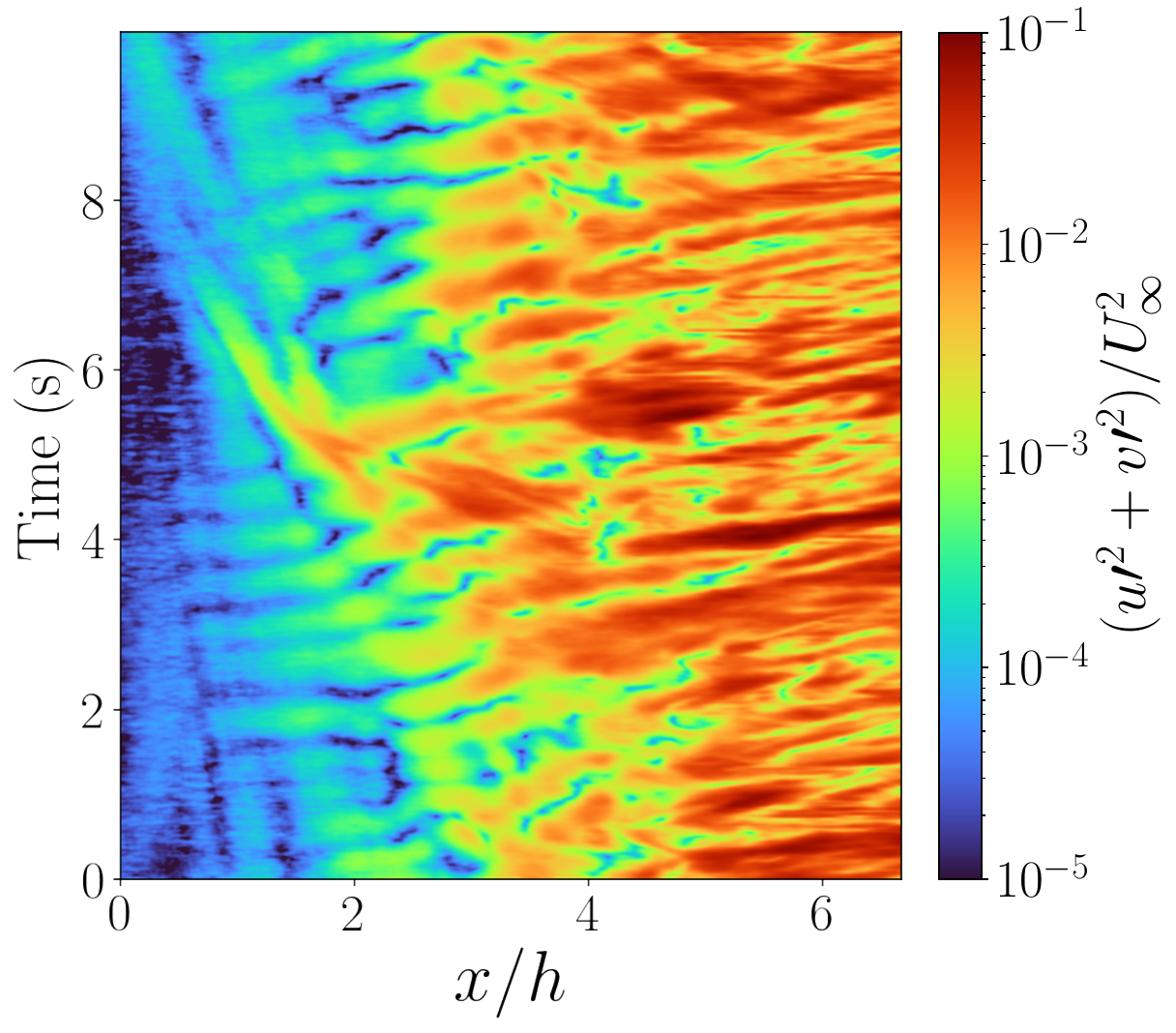}
        \caption{}
        \label{fig:space_time_fke}    
    \end{subfigure}
    \caption{\textcolor{black}{ Space--time diagrams of a horizontal line taken at $y/h = 0.5$ in the 2D velocity fields over the time-window of the recorded extreme event. (a) Spatio-temporal evolution of the vorticity. (b) Spatio-temporal evolution of the fluctuation of kinetic energy. In both cases, the strong ejection in the upstream direction is clearly visible. }}
    \label{fig:space_time}
\end{figure}

\textcolor{black}{To complement the description of the dynamics around the detected event, we present in Fig.~\ref{fig:space_time} space--time diagrams of vorticity and $FKE$ extracted from a longitudinal profile at $y/h=0.5$, downstream of the BFS. This representation is useful to visualize the full time history of the detected event. In this representation, structures with a positive slope are advected downstream of the BFS, while structures with negative slopes move upstream, toward the BFS. Both diagrams show the presence of regular structures being advected downstream in the second half of the domain, from $x/h \approx 4$ to $x/h \approx 7$. In the first half of the domain, a clear bursting in the upstream direction, corresponding the detected event, can be observed. 
The space--time evolution identified here will be connected to global metrics (space-averaged enstrophy and fluctuating energy) in Sec.~\ref{sec:discussion}, which peak concurrently with the onset of the upstream injection.}

%% file: Discussion.tex
\section{Discussion\label{sec:discussion}}

\begin{figure}[h]
\begin{subfigure}{0.48\textwidth}
    \centering
    \includegraphics[width=\linewidth]{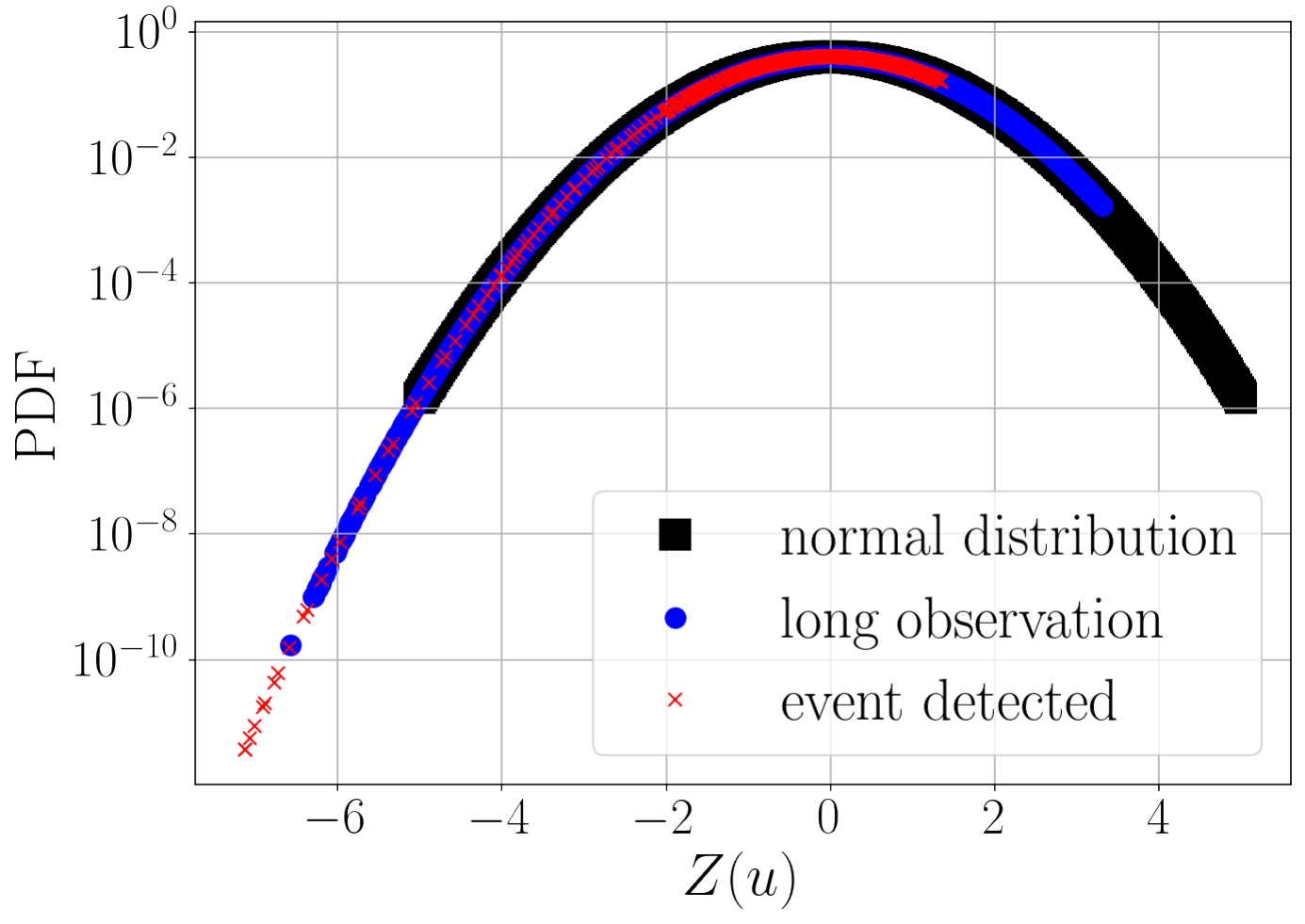}
    \caption{}
    \label{fig:distro_u}
\end{subfigure}
\begin{subfigure}{0.48\textwidth}
    \centering
    \includegraphics[width=\linewidth]{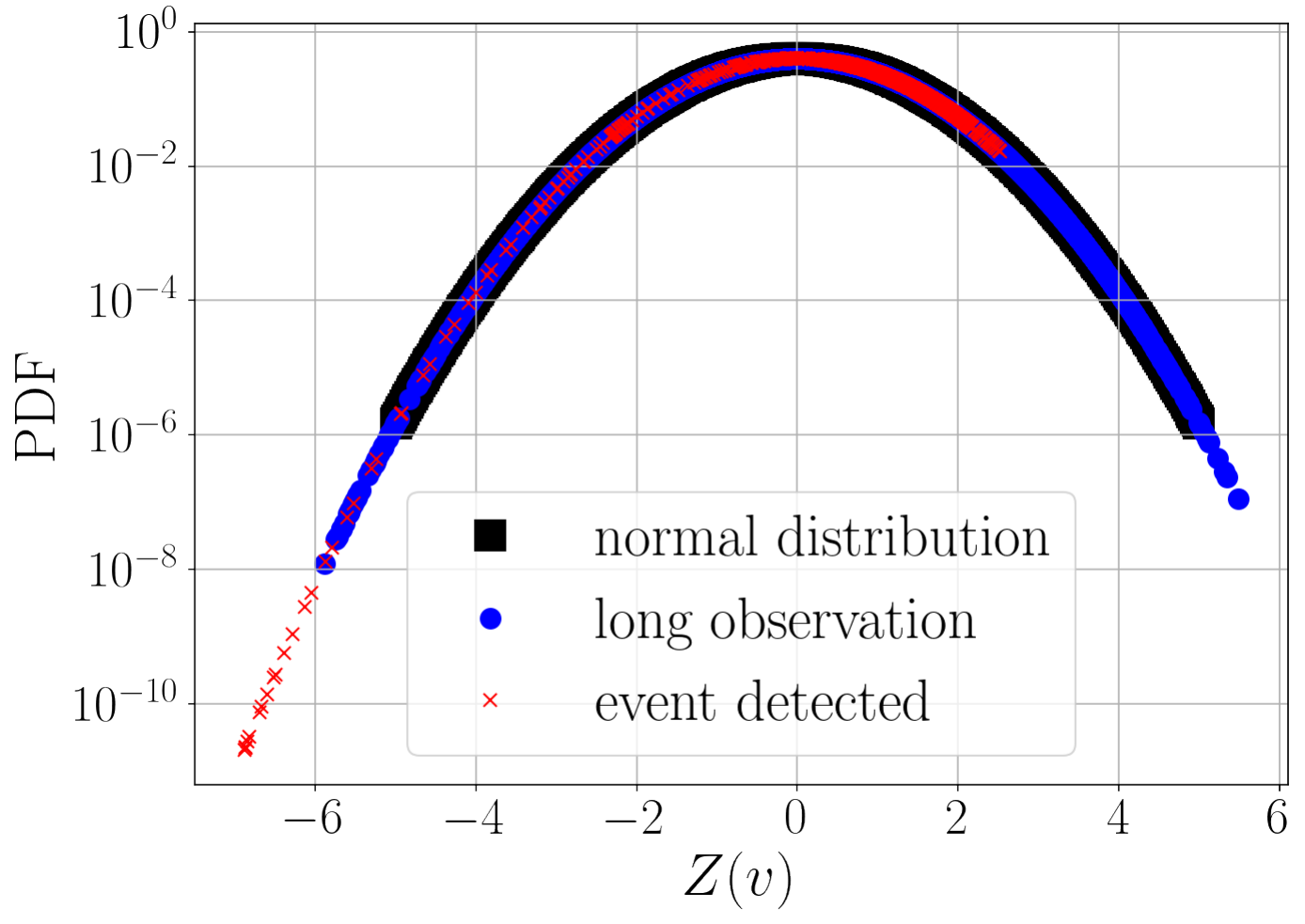}
    \caption{}
    \label{fig:distro_v}
\end{subfigure}
\caption{PDF comparison of the time series of the probes from the initial long observation, the detected event and a normal distribution. a) $u$ component. b) $v$ component}
\label{fig:distros}
\end{figure}

The detected event exhibits clear signatures of intermittency and bursting that can be characterized statistically. Time series from the probe $n^{o}$1 ($(x,y)=(2h,h/2)$) displays marked non-normal behavior. Figure~\ref{fig:distros} compares the PDFs of both velocity components with a standard normal distribution, for both the long observation and the detected event. For the streamwise component ($u$) (Fig.~\ref{fig:distro_u}) the long observation is non-normal, with a skewness $ \gamma_1 =-0.725$ and excess kurtosis $ \kappa = 1.164$, consistent with a negatively skewed distribution with a heavy tail toward $Z(u)<-6$. During the detected event, the non-Gaussianity is amplified (skewness $\gamma_1=-2.458$, excess kurtosis $\kappa = 8.261$). The wall-normal component ($v$), Fig.~\ref{fig:distro_v}, is closer to normal in the long observation (skewness $\gamma_1=0.0422$, excess kurtosis $\kappa = 0.494$), but becomes strongly non-Gaussian during the event ($\gamma_1=-2.356$, $\kappa = 6.519$).

\begin{figure}[h]
\begin{subfigure}{0.48\textwidth}
    \centering
    \includegraphics[width=\linewidth]{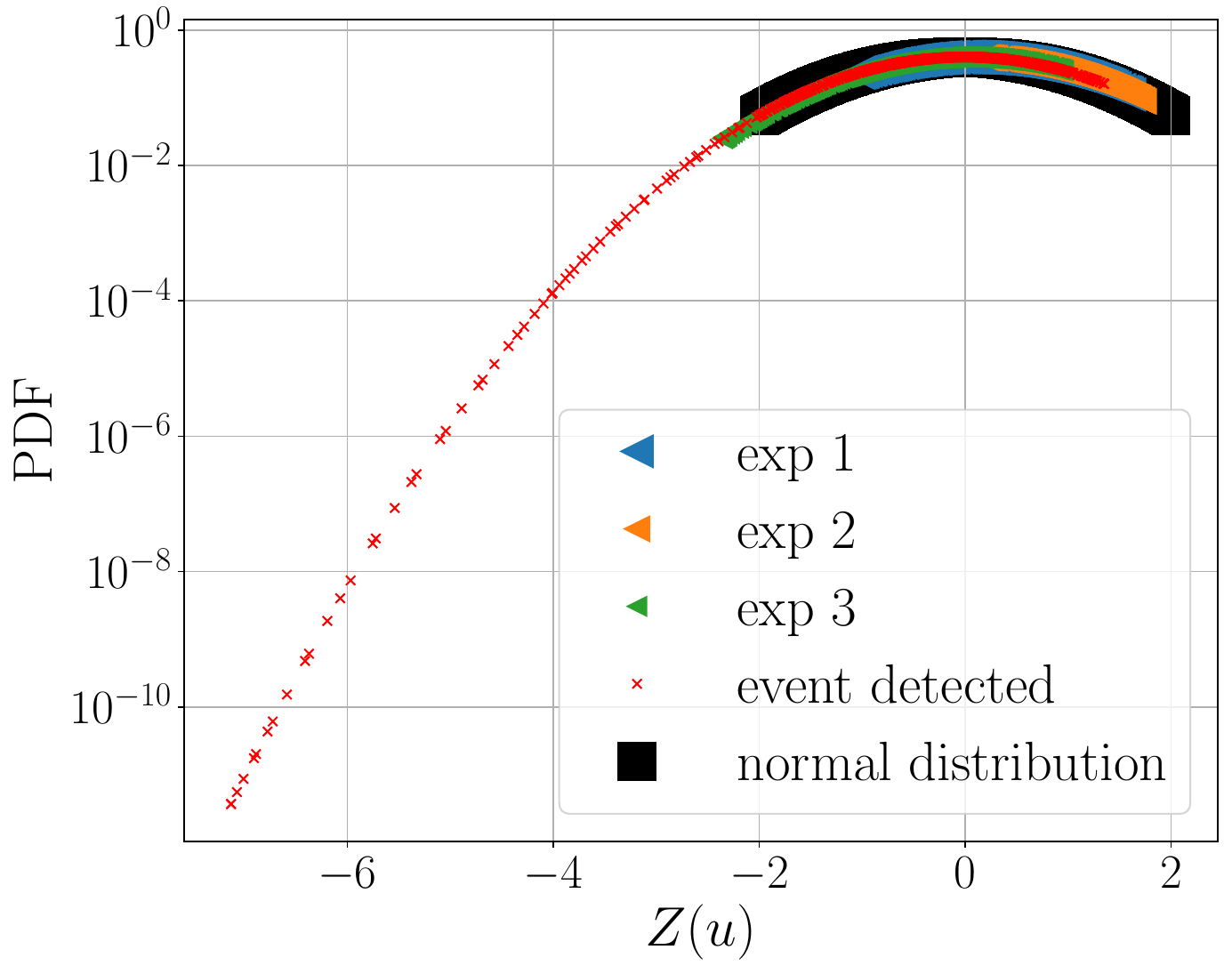}
    \caption{}
    \label{fig:distro_u_det}
\end{subfigure}
\begin{subfigure}{0.48\textwidth}
    \centering
    \includegraphics[width=\linewidth]{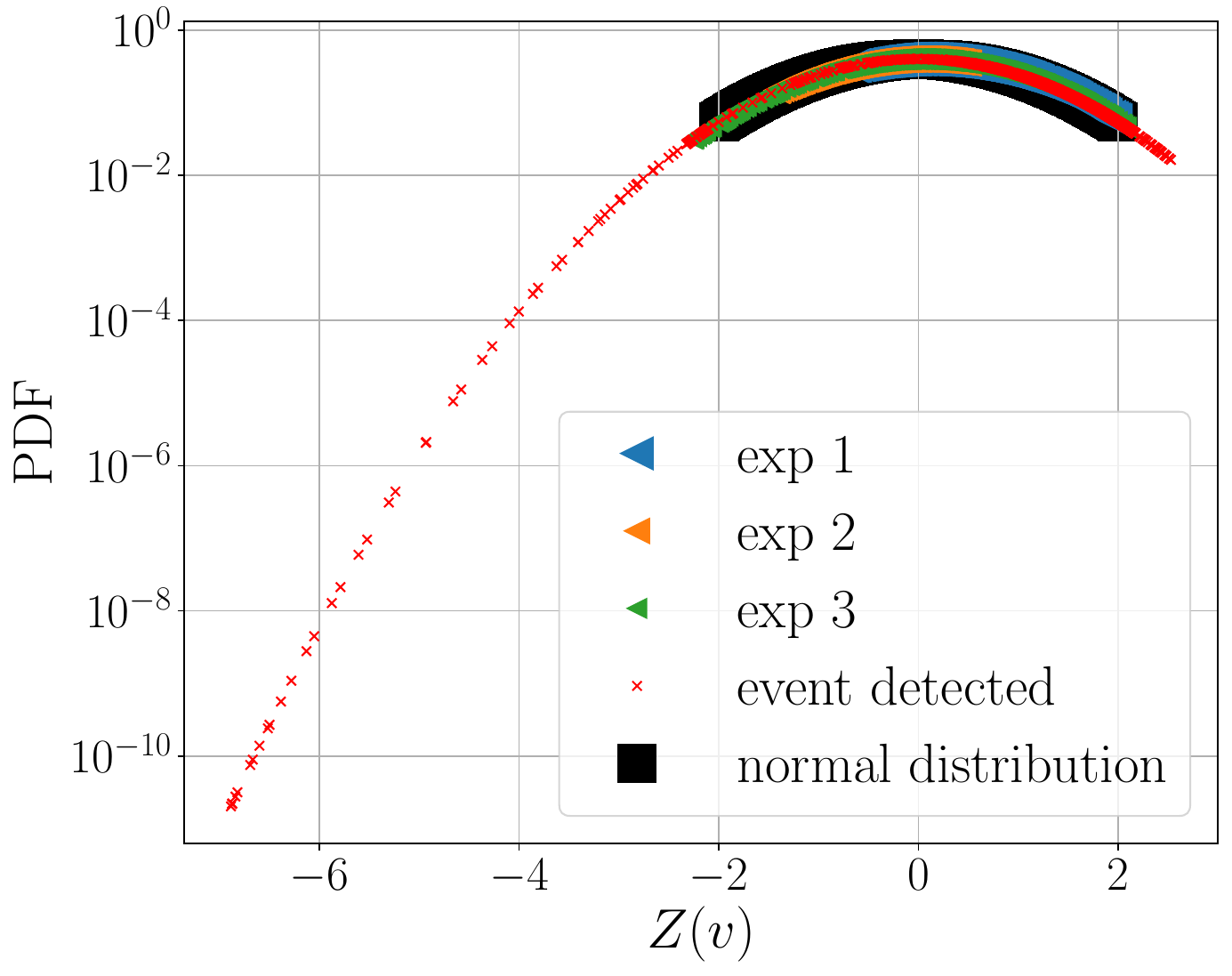}
    \caption{}
    \label{fig:distro_v_det}
\end{subfigure}
\caption{PDF comparison of the probes time series from the detected event, the baseline  and a normal distribution. a) $u$ component. b) $v$ component}
\label{fig:distros_det}
\end{figure}

This non-normal behavior is not unique to the detected event. Figure~\ref{fig:distros_det} shows the PDFs of $u$ and $v$ for the detected event and for the three random baseline samples, compared against a standard normal distribution over $-2\le Z\le 2$. The baseline datasets already display slight departures from normality, which is confirmed by the statistics in Table~\ref{tab:statistics}: small skewness and \emph{negative} excess kurtosis (platykurtic) are observed for the baselines. The detected event, however, substantially accentuates the non-normality, with strong negative skewness and large positive excess kurtosis. This is consistent with intermittent, bursty dynamics \citep{She1991IntermittencyAN,Li2005OriginOn} and indicates that the flow is driven into a distinct statistical state during the event.

\begin{table}[h]
    \centering
    \begin{tabular}{|c|c|c|c|}
    \hline
         \textbf{Exp}& \textbf{Component}&\textbf{skewness}&\textbf{Excess kurtosis} \\
         \hline
         1 &  $u$ & -0.6134 & -0.1766 \\
        \hline
        1 & $v$ & 0.1656 & -0.9184 \\
        \hline
        2 &  $u$ & -0.3884 & -0.3580 \\
        \hline
        2 & $v$ & 0.1296 & -0.8938 \\
        \hline
        3 &  $u$ & -0.1955 & -0.2523 \\
        \hline
        3 & $v$ & -0.4102 & -0.2041 \\
        \hline
        Event & $u$ & -2.4583 & 8.2605  \\
        \hline
        Event & $v$ & -2.3565 & 6.5189 \\
        \hline
    \end{tabular}
    \caption{Skewness and excess kurtosis of time series at probe $n^{o}$1 ($(x,y)=(2h,h/2)$) for the baseline and the detected event. The numbers in the column \textbf{Exp} mark the baseline.}
    \label{tab:statistics}
\end{table}

\begin{figure}[h]
    \centering
    \includegraphics[width=0.75\linewidth]{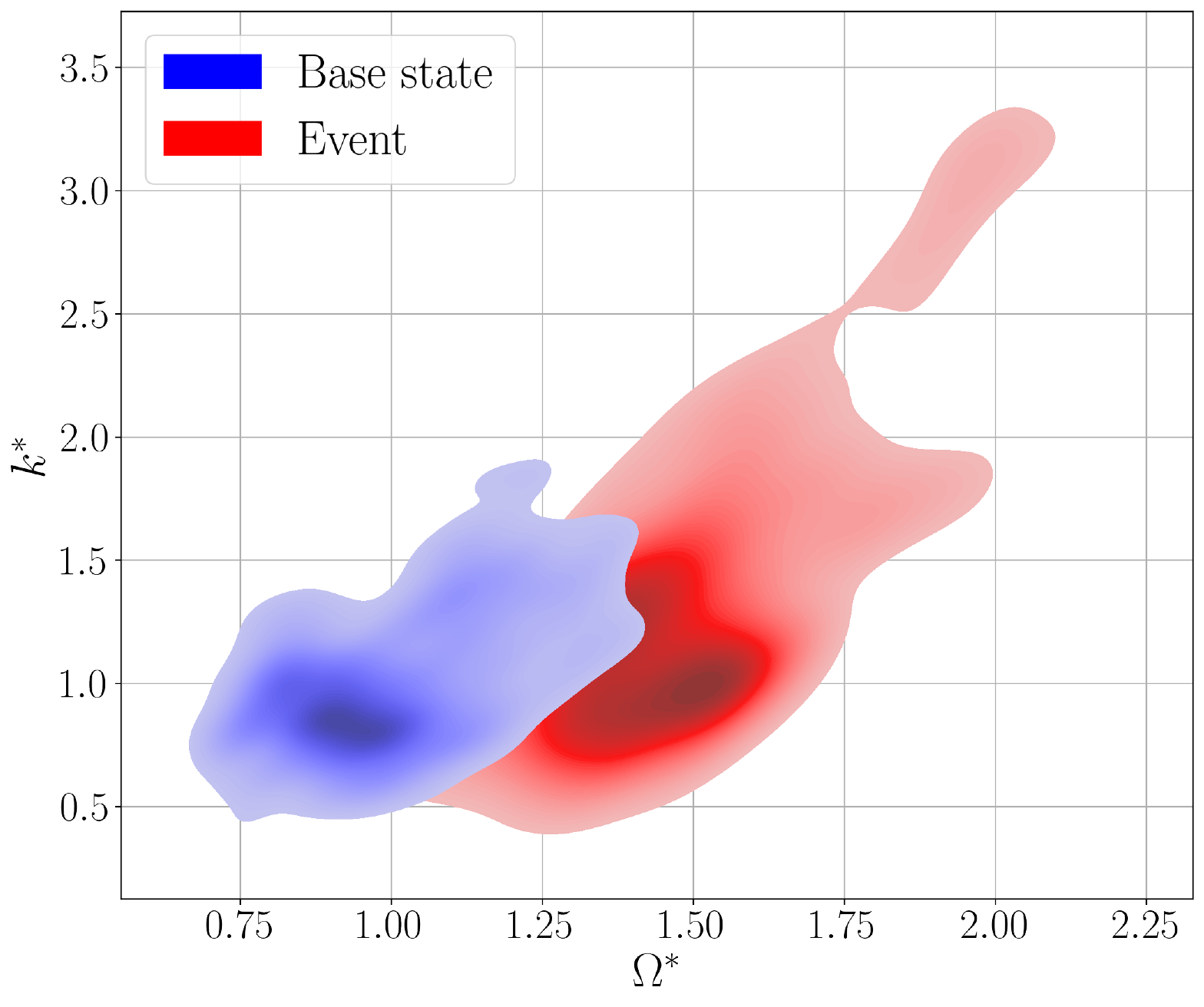}
    \caption{Joint PDF of space averaged fluctuation of kinetic energy and enstrophy for the base state in blue and the detected event in red. both quantities are normalized by the mean value of the base state. }
    \label{fig:KDE}
\end{figure}

To connect these local statistics to global field dynamics, the space-averaged fluctuating kinetic energy $k'$ and enstrophy $\Omega$ from the fluctuating fields are computed using the following definitions:

\begin{gather}
    k'(t) = \frac{1}{2}\,\big\langle u'(x,y,t)^2 + v'(x,y,t)^2 \big\rangle_{x,y},
    \qquad
    k^*(t) = \frac{k'(t)}{\big\langle k'(t) \big\rangle_{\text{base}}},
    \label{eq:tke}
\end{gather}

\begin{gather}
    \omega'(x,y,t) = \frac{\partial v'(x,y,t)}{\partial x} - \frac{\partial u'(x,y,t)}{\partial y},
    \qquad
    \Omega(t) = \frac{1}{2}\,\big\langle \omega'(x,y,t)^2 \big\rangle_{x,y},
    \qquad
    \Omega^*(t) = \frac{\Omega(t)}{\big\langle \Omega(t) \big\rangle_{\text{base}}}.
    \label{eq:om}
\end{gather}

Here $\langle \cdot \rangle_{x,y}$ denotes spatial averaging over the field of view. The baseline (subscript ``base'') is the ensemble mean constructed from the three baseline acquisitions; fluctuations $u',v'$ are defined relative to this baseline.

Figure~\ref{fig:KDE} shows the joint PDF of $(k^*,\Omega^*)$ for the baseline and for the detected event. The baseline spans up to $\sim 1.5$ times its mean along the enstrophy axis and up to $\sim 2$ along the energy axis, with highest density near $(0.9,0.75)$. During the event, the distribution is shifted toward larger values in both coordinates, indicating a \emph{simultaneous} amplification of rotational activity and fluctuating energy. The region of highest density is clearly displaced from the baseline cluster, consistent with a distinct, more energetic/vortical state.

\begin{figure}[h]
    \centering
    \includegraphics[width=\linewidth]{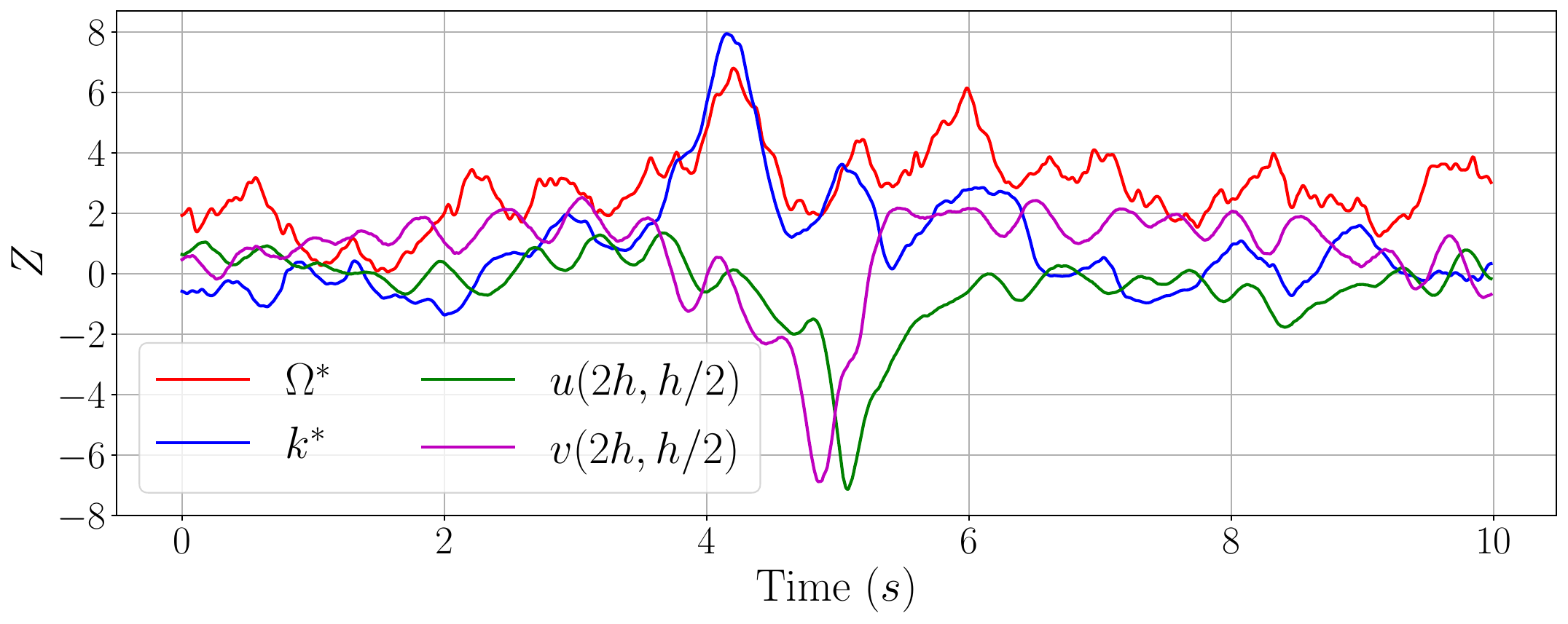}
    \caption{Time series of global and local quantities, expressed in terms of their $Z$ score with respect to the baseline.}
    \label{fig:Z_global_local}
\end{figure}

These trends are corroborated in Fig.~\ref{fig:Z_global_local}, which plots the Z-scores of $k^*$, $\Omega^*$, together with the local probe of velocity signals for the detected event. The largest deviations of enstrophy and energy occur \emph{simultaneously}, and precede the large negative excursions in the local velocity probes. This timing is consistent with the scenario described in Sec.~\ref{sec:event_detected}: first, a surge in vortical activity and fluctuation energy (collapse of the merged KH vortex), followed by upstream mass injection into the recirculation region that reaches $x/h\approx2$ and produces the extreme local deviations. The peaks of $k^*$ and $\Omega^*$ occur at $t\approx 4.2~\mathrm{s}$ (see Fig.~\ref{fig:VF_t4}), coinciding with the onset of \textcolor{black}{the} upstream-wise bursting jet.

Altogether, the statistical signatures (skewness/kurtosis), the joint $(k^*,\Omega^*)$ shift, and the temporal ordering of global and local deviations support the interpretation of the detected event as a rare, intermittent transition involving a bursting jet toward the BFS.

%% file: conclusions.tex
\section{Conclusions\label{sec:conclusions}}

We have demonstrated how Live Optical Flow Velocimetry (L-OFV) can be used for long-duration monitoring of a backward-facing step (BFS) flow, enabling real-time detection and capture of rare, extreme events. An initial statistical analysis of local probe signals established data-driven thresholds based on the heavy tails of the observed PDFs. In practice, detection relied on large negative excursions of the probes Z-scores (e.g., $u:~Z<-6$, $v:~Z<-5$ at $(x,y)=(2h,h/2)$), which triggered the recording of the velocity fields before and after the event.

A single, unique event was captured and analyzed in detail. The velocity fields revealed a jet-like fluid motion towards the BFS, into the recirculation region. It seems initiated by the collapse of a merged Kelvin–Helmholtz vortex and subsequently propelled by counter-rotating vortices between the shear layer and near the wall. The injected flow ultimately rolled up into a slowly turning vortex that persisted near $x/h\approx 2$. The temporal sequence of global and local indicators was consistent with this mechanism: space-averaged fluctuating kinetic energy and enstrophy peaked first (at $t\approx 4.2~\mathrm{s}$), followed by the large negative excursions in the local probe signals as the upstream injection reached the probe location.

Statistically, the event departed strongly from Normal behavior. The probe signals exhibited heavy tails and strong negative skewness with large positive excess kurtosis, together with pronounced excursions in the $(u,v)$ phase space. At the field level, the joint PDF of normalized mean fluctuating kinetic energy and mean enstrophy $(k^*,\Omega^*)$ shifted towards higher values during the event, indicating a simultaneous amplification of turbulent fluctuations and rotational activity. These signatures are consistent with an intermittent, burst-like transition of the separated shear layer.

The present study is limited by the observation of a single event within a long monitoring campaign. It precludes robust estimation of occurrence rates, which would require many successive monitoring experiments. Nevertheless, it demonstrates the feasibility of new types of experiments dedicated to the search for rare and extreme events in complex shear flows using L-OFV.

Future work should (i) extend monitoring durations to improve event statistics, (ii) explore a broader range of Reynolds numbers, and (iii) refine detection criteria by combining local thresholds with simultaneous peaks of global indicators (e.g., $(k^*,\Omega^*)$) or extreme-value-theory-based metrics. Such efforts will help quantify the incidence and mechanisms of extreme events in separated flows and further establish real-time velocimetry as a powerful tool for experimental rare-event research.